\documentclass[10pt,letterpaper]{article}
\usepackage[top=0.85in,left=2.75in,footskip=0.75in]{geometry}

\usepackage{float}
\usepackage{amsfonts}
\usepackage{mathtools}
\usepackage{xr-hyper}
\usepackage{gensymb}

\usepackage{amsmath,amssymb}

\usepackage{changepage}

\usepackage[utf8x]{inputenc}

\usepackage{textcomp,marvosym}

\usepackage{cite}

\usepackage{nameref,hyperref}

\usepackage[right]{lineno}

\usepackage{microtype}
\DisableLigatures[f]{encoding = *, family = * }

\usepackage[table]{xcolor}

\usepackage{array}

\newcolumntype{+}{!{\vrule width 2pt}}

\newlength\savedwidth

\usepackage{setspace} 
\raggedright
\usepackage[aboveskip=1pt,labelfont=bf,labelsep=period,justification=raggedright,singlelinecheck=off]{caption}

\makeatletter
\renewcommand{\@biblabel}[1]{\quad#1.}
\makeatother

\usepackage{lastpage,fancyhdr,graphicx}
\usepackage{epstopdf}
\fancyheadoffset[L]{2.25in}
\fancyfootoffset[L]{2.25in}
\newcommand\customspace{\vspace{4pt}}

\DeclareMathOperator*{\argmin}{argmin}

\newcommand{\VMD}{\text{LD}}
\newcommand{\LD}{\text{LD}}

\newcommand{\LM}{\text{LM}}

\newcommand{\R}{\mathbb{R}}

\newcommand{\rd}{\mathrm{d}}

\makeatletter
\renewcommand*{\fps@figure}{t}  %Switch between t and H
\makeatother
\newif\ifshow
\showtrue       % \showtrue for showing figures
\newcommand{\showFigure}[1]{\ifshow #1\fi}
\begin{document}
\vspace*{0.2in}

\begin{flushleft}
{\Large
\textbf{Analysis of protrusion dynamics in amoeboid cell motility by means of regularized contour flows
% Please use "sentence case" for title and headings (capitalize only the first word in a title (or heading), the first word in a subtitle (or subheading), and any proper nouns).
}}
\newline
\\
Daniel Schindler\textsuperscript{1},
Ted Moldenhawer\textsuperscript{2},
Maike Stange\textsuperscript{2},
Valentino Lepro\textsuperscript{2,3},\\
Carsten Beta\textsuperscript{2},
Matthias Holschneider\textsuperscript{1},
Wilhelm Huisinga\textsuperscript{1*}
\\
\bigskip

$^1$ Institute of Mathematics, University of Potsdam, Potsdam, Germany,\\
$^2$ Institute of Physics and Astronomy, University of Potsdam, Potsdam, Germany\\
$^3$ Max Planck Institute of Colloids and Interfaces, Potsdam, Germany
\bigskip

% Insert additional author notes using the symbols described below. Insert symbol callouts after author names as necessary.
% 
% Remove or comment out the author notes below if they aren't used.
%
% Primary Equal Contribution Note
%\Yinyang These authors contributed equally to this work.

% Additional Equal Contribution Note
% Also use this double-dagger symbol for special authorship notes, such as senior authorship.
%\ddag These authors also contributed equally to this work.

%\textcurrency Current Address: Dept/Program/Center, Institution Name, City, State, Country % change symbol to "\textcurrency a" if more than one current address note
% \textcurrency b Insert second current address 
% \textcurrency c Insert third current address


% Group/Consortium Author Note
%\textpilcrow Membership list can be found in the Acknowledgments section.

% Use the asterisk to denote corresponding authorship and provide email address in note below.
* huisinga@uni-potsdam.de

\end{flushleft}
\section*{Abstract}
% setting the scene
Amoeboid cell motility is essential for a wide range of biological processes including wound healing, embryonic morphogenesis, and cancer metastasis. 
% what's the problem?
It relies on complex dynamical patterns of cell shape changes that pose long-standing challenges to mathematical modeling
and raise a need for automated and reproducible approaches to extract quantitative morphological features from image sequences.
% what do we do and how?
Here, we introduce a theoretical framework and a computational method for obtaining smooth representations of the spatiotemporal contour dynamics from stacks of segmented microscopy images.
Based on a Gaussian process regression
we propose a one-parameter family of regularized contour flows that allows us to continuously track reference points (virtual markers) between successive cell contours.
% what's the benefit?
We use this approach to define a coordinate system on the moving cell boundary and to represent different local geometric quantities in this frame of reference.
In particular, we introduce the local marker dispersion as a measure to identify localized membrane expansions and provide a fully automated way to extract the properties of such expansions, including their area and growth time. The methods are available as an open-source software package called \texttt{AmoePy}, a Python-based toolbox for analyzing amoeboid cell motility (based on time-lapse microscopy data), including a graphical user interface and detailed documentation.
% application
Due to the mathematical rigor of our framework, we envision it to be of use for the development of novel cell motility models.
We mainly use experimental data of the social amoeba {\it Dictyostelium discoideum} to illustrate and validate our approach. 

% ======================================================================================

% Please keep the Author Summary between 150 and 200 words
% Use first person. PLOS ONE authors please skip this step. 
% Author Summary not valid for PLOS ONE submissions.

\section*{Author summary}
% setting the scene
Amoeboid motion is a crawling-like cell migration that plays an important key role in multiple biological processes such as wound healing and cancer metastasis.
This type of cell motility results from expanding and simultaneously contracting parts of the cell membrane.
% what's the problem?
From fluorescence images, we obtain a sequence of points, representing the cell membrane, for each time step.
By using regression analysis on these sequences, we derive smooth representations, so-called contours, of the membrane.
Since the number of measurements is discrete and often limited, the question is raised of how to link consecutive contours with each other.
%The number of measurements is limited by the level of radiation, which affects and eventually even kills the cell. This also limits the number of contours and raises the question how to link consecutive contours with each other.
% what do we do and how?
In this work, we present a novel mathematical framework in which these links are described by regularized flows allowing a certain degree of concentration or stretching of neighboring reference points on the same contour.
% what's the benefit?
This stretching rate, the so-called local dispersion, is used to identify expansions and contractions of the cell membrane providing a fully automated way of extracting properties of these cell shape changes.
% application
We applied our methods to time-lapse microscopy data of the social amoeba {\it Dictyostelium discoideum}.

%\linenumbers

%%%%%%%%%%%%%%%%%%%%%%%%%%%%%%%%%%%%%%%%%%%%%%%%%%%%%%%%
%%%%%%%%%%%%%%%%%%%%%%%%%%%%%%%%%%%%%%%%%%%%%%%%%%%%%%%%
% Introduction
%Use "Eq" instead of "Equation" for equation citations.
\section*{Introduction}\label{sec:Intro}
Amoeboid motion is one of the most widespread forms of cell motility in the living world~\cite{bray_cell_2000}.
It plays a key role in many essential functions of the human body, such as responses of the immune system~\cite{germain_decade_2012} or the healing of injured tissue~\cite{shaw_wound_2009}.
Its medical relevance also extends to the field of cancer research, as metastatic tumor cells rely on amoeboid motility to invade the surrounding tissue~\cite{condeelis_great_2005}.
Amoeboid locomotion is based on dynamical changes of the cell shape.
Specifically, localized protrusions of the cell membrane, often called pseudopodia, that are extended in the direction of motion, are generally seen as the basic morphological entities that drive amoeboid movement~\cite{haastert_how_2011}.
Together with membrane contraction at the back of the cell body, their extension results in a displacement of the center of mass of the cell. 
This requires, besides the coordinated pattern of protrusion and retraction, also the formation and rupture of adhesive contacts to a substrate or to a surrounding extracellular matrix~\cite{lauffenburger_cell_1996}.

The mechanical forces that drive the shape changes of amoeboid cells are generated by the actin cytoskeleton, a dynamic filament meshwork at the inner face of the cell membrane~\cite{blanchoin_actin_2014}. 
The main building blocks of this network are actin filaments that are subject to a constant turnover by polymerization and depolymerization, resulting in a continuous rapid reorganization of the network structure.
This process is assisted by a host of auxiliary cytoskeletal proteins that initiate the nucleation, capping, or severing of actin filaments as well as filament bundling, branching, and membrane cross-linking.
The cytoskeletal machinery is orchestrated by biochemical signaling pathways that coordinate the spatiotemporal patterns of activity in the actin system across the cell cortex~\cite{ridley_cell_2003}. 
These upstream signaling pathways also provide a link to membrane receptor signals, so that cells may react to gradients of extracellular cues by moving directionally towards a chemical source---a process commonly referred to as chemotaxis~\cite{haastert_chemotaxis_2004}.

The mathematical modeling of amoeboid motion is a long-standing challenge that has been addressed at different levels of complexity, ranging from random walk models for the center of mass of the cell~\cite{selmeczi_cell_2008,bodeker_quantitative_2010,makarava_quantifying_2014} up to detailed high-dimensional models for the intracellular signaling activity~\cite{khamviwath_continuum_2013}.
Typically, current models focus on selected mechanistic aspects of amoeboid motility and describe, for example, actin dynamics, cell-to-cell variability, or the switching between different migratory modes in more detail, see for example~\cite{dreher_spiral_2014,alonso_modeling_2018,cao_plasticity_2019}.
Also, first attempts have been made to incorporate several key components, such as dynamic signaling patterns, polarity formation, and cytoskeletal activity, in a modular approach~\cite{devreotes_excitable_2017}.
These models, however, remain qualitative and their comparison with experimental data oftentimes relies on visual inspection.
As the entire biological process involves many hundreds of interacting proteins and signaling molecules, with many mechanistic details yet unknown, a quantitative model that includes the full molecular details remains out of reach.

To advance the mathematical description of amoeboid motility, we envision that current efforts of mechanism-based modeling are complemented by a more systematic, data-driven approach. 
This requires a mathematical framework that allows us to systematically develop a quantitative model based on experimental data. 
Such a framework should rely on observables that encode the key characteristics of amoeboid motility and, at the same time, are readily accessible experimentally.
Trajectories of the center-of-mass of the cell can be easily recorded in large amounts from low-resolution bright-field microscopy data but reflect only very limited, integral information on the entire process.
The intracellular signaling pathways and the cytoskeletal mechanisms, on the other hand, are difficult to access and knowledge on this part of the system remains highly incomplete.
We therefore concentrate on the cell shape as the central reference quantity.
The cell shape is fully accessible by standard microscopy techniques and can be easily recorded with sufficient spatial and temporal resolution.
Moreover, its dynamic evolution implicitly reflects the intracellular processes and determines the center-of-mass trajectory of the cell.

In our long-term quest for a quantitative, data-driven model of amoeboid motility, several steps are required:
First, the development of a mathematical framework for the description of experimentally observed shape dynamics;
second, the design of a model of the contour dynamics that predicts realistic shape evolutions; and
finally, the incorporation of mechanistic information on key intracellular processes as the driving determinants of the contour dynamics. 
These become accessible by imaging of fluorescently tagged fusion proteins and by more advanced methods, such as knock-sideways and optogenetic approaches. 
Here, existing mechanistic models may provide a useful basis and might merge into a joint modeling concept.

In this article, we concentrate on the first aspect of the above agenda.
We provide a mathematically well-defined approach that allows for a detailed analysis of the complex, multifaceted contour dynamics of amoeboid cells. 
A key ingredient is the concept of regularized flows between contours that define an evolution of virtual markers in time. Using contour flows, we define a coordinate system on the evolving contours (strongly regularized case) and approximate local quantities of interest (weakly regularized case).
While the strongly regularized flow is used to define trajectories of virtual markers over the entire time span, the weakly regularized flow is used to obtain information on local membrane changes, which is mapped subsequently onto the global flow representing the coordinate system. This separation between global and local flows is an essential feature of our approach.

Established approaches to describe virtual markers on an evolving contour most commonly include level set methods (LSM), e.g.,
in~\cite{wolgemuth_2010}, where the cell contour dynamics was additionally decomposed into a translation of the entire cell and the deformation of its contours. In~\cite{machacek_morphodynamic_2006}, the LSM was compared to a mechanistic spring model penalizing a dense concentration of virtual markers. While the mechanistic model provides better computation times than the LSM, it shows limitations regarding topological mapping violations during strong shape deformations.
Alternatively, electrostatic field equations were successfully employed, tackling mapping violation issues while providing better computation times \cite{tyson_high_2010}. However, it does not provide trajectories of virtual markers over the entire time interval in its current form.
In~\cite{driscoll_cell_2012}, a mapping was chosen which minimizes the sum of squared distances between virtual markers (so-called mean squared displacement), while enforcing an evenly-spaced distribution of virtual markers. Our work aims at (i) presenting a theoretical framework for the analysis of marker dynamics and (ii) combining the respective merits of the above approaches into a single method.

In the spirit of previous approaches, we rely on the widely used concept of kymographs to graphically represent the space-time dynamics of different geometric quantities, such as the speed of membrane displacement or the local curvature, along the cell contour.
In this context, we propose a novel criterion based on a single dynamic quantity for defining membrane expansions/contractions.
Prior approaches to identify membrane protrusions relied on the simultaneous matching of multiple criteria, such as critical values of curvature, protrusion speed, and pseudopod lifetime~\cite{bosgraaf_quimp3_2010} or identified protrusion events as points in time only~\cite{driscoll_cell_2012}, without providing major protrusion properties, such as area growth rate, shape, and others.
In \cite{cooper_2012}, pseudopods were detected by using a hierarchical clustering algorithm in which individual membrane extensions are connected based on direction and continuity in space and time.
Furthermore, image skeletonization on contours was used to identify and characterize pseudopods in an automated way \cite{xiong_automated_2010}.
Unfortunately, all existing approaches have in common an undesirable blending of the protrusion-defining criteria with their numerical implementation.
This makes it difficult to discern biological effects from numerical artifacts.
We instead first define our criterion in mathematical terms and only subsequently implement it numerically, allowing us to control numerical errors and avoid artifacts.

The methodology of this article is implemented in \texttt{AmoePy}, a Python-based toolbox for analyzing and simulating amoeboid cell motility~\cite{amoepy_2020}. Furthermore, \texttt{AmoePy} features multiple data handling and analysis tools, an easy-to-use graphical user interface, and data files containing multiple experimental cell tracks in addition to simple artificial tracks which were used to additionally validate the algorithm.

The article is organized as follows.
We first develop the mathematical framework and introduce the concept of regularized contour flows.
We then illustrate our approach in applications to experimental recordings of the social amoeba {\it Dictyostelium discoideum}, a widely used model organism for the study of amoeboid motility.
Finally, based on the results of our analysis, we illustrate in the Discussion section how we envision the next steps towards a quantitative, data-driven model of amoeboid motility, based on a point process of the protrusive activity.

%\begin{itemize}
%    \item \red{General problem/Context}
%    \item \red{Specific Problem}
%    \item \red{Background}
%    \item \red{Approach}
%\end{itemize}

%%%%%%%%%%%%%%%%%%%%%%%%%%%%%%%%%%%%%%%%%%%%%%%%%%%%%%%%
%%%%%%%%%%%%%%%%%%%%%%%%%%%%%%%%%%%%%%%%%%%%%%%%%%%%%%%%
% ===================================================================================
\section*{Methods}\label{sec:Methods}
% ===================================================================================

%--------------------------------------------------------------------------------------------------------------------------------------------------
\subsection*{Data acquisition and image segmentation}
%--------------------------------------------------------------------------------------------------------------------------------------------------

Experiments were performed with AX2 cells of the social amoeba {\it Dictyostelium discoideum}. 
As a marker for filamentous actin, cells expressed fluorescently tagged Lifeact (C-terminally fused with mRFP, plasmid kindly provided by Igor Weber, Zagreb, HR).
They were grown at $20\degree$C in liquid culture flasks containing HL5 medium including glucose (Formedium, Hunstanton, GB) and 10~$\mu$g/ml G-418 sulfate (Cayman Chemical Company, US) as a selection marker.
Before each experiment, cells were harvested from culture flasks and grown in 25~ml overnight shaking cultures at 180 rpm under otherwise identical conditions.
Afterward, nutrients were removed by centrifugation and washing of the cell pellet with S{\o}rensen phosphate buffer at pH~6 (14.7~mM KH$_2$PO$_4$, Merck, Darmstadt, DE; 2~mM Na$_2$HPO$_4 \times\!$ H$_2$O, Merck, Darmstadt, DE).
Then, cells were resuspended in fresh buffer and droplets were formed in a Petri dish to initiate the streaming processes.

For image acquisition cells were transferred after 5~hours to a glass bottom dish (Fluorodish, ibidi GmbH, Martinsried, DE).
During imaging, they were kept in S{\o}rensen phosphate buffer at $20\degree$C.
Images were taken with a Zeiss LSM780 laser scanning confocal microscope (Carl Zeiss AG, Oberkochen, DE) at a frame rate of one image per second, using a $63\times$ or $40\times$ oil immersion objective.
Fluorophores were excited at 651~nm and emission was recorded between 562~nm and 704~nm.
For details see also~\cite{alonso_modeling_2018}, where the same data set was used in a different context.

Fluorescence image (8-bit gray scale) were segmented using a modified version of the active contour (snake) algorithm described in~\cite{driscoll_local_2011,driscoll_cell_2012}. 
Based on this algorithm, we parameterized the cell boundary in each frame by a closed string of $M=400$ equidistant nodes.
As frames were taken at discrete time points $t_0,\ldots,t_{K-1}$ with equal time difference $\delta t = t_{k+1}-t_k = 1$~sec, 
we denoted the discrete representation of the cell contour in a given frame at time $t_k$ as $\gamma_k$ with supporting points
\begin{equation} \label{eq:data}
    (x_{k,0},y_{k,0}),\ldots, (x_{k,M-1},y_{k,M-1}) \in\R^2.
\end{equation}
In the result section, the data sets consists of $K=500$ to $1000$ time frames. For later reference, we set $t_0=0$ and $t_{K-1}=T$. 

%--------------------------------------------------------------------------------------------------------------------------------------------------
\subsection*{Estimate of contour dynamics}
%--------------------------------------------------------------------------------------------------------------------------------------------------

We used a real-valued smoothing spline for the $x$ and $y$ coordinates based on Gaussian process regression (GPR) using a Poisson kernel, for details see \nameref{s:appendix1}.
This yielded a parametrization of the contour $\Gamma_k$ at time 
frame $t_k = k\cdot \delta t$
\begin{equation}\label{eq:contourmaps}
    \Phi_k : [0, 2\pi) \owns\theta \mapsto ( x(\theta), y(\theta))  \in \mathbb{R}^2
\end{equation}
in terms of a finite sum of smooth kernels (see e.g. \cite{Rasmussen2006,Bousquet2004} for details)
\begin{equation*}
x(\theta) = \sum_{m=0}^{M-1} c_m P(\theta_m-\theta),\qquad  y(\theta) = \sum_{m=0}^{M-1} d_m P(\theta_m-\theta)
\end{equation*}
where $P$ is a suitably scaled Poisson kernel. Support points were chosen
to correspond to normalized secant length along the contour 
\begin{align*}
\theta_{k,m}& = \frac{2\pi \sum_{i=0}^m d_{k,i}}{ \sum_{i=0}^{M-1} d_{k,i}},\quad  d_{k,0}= 0, \quad d_{k,i} = \left((x_{k,i}-x_{k,i-1})^2 + (y_{k,i}-y_{k,i-1})^2\right)^{1/2}
\end{align*}
for $m=0,\ldots,M-1$ and $k=0,\ldots,K-1$, see \nameref{s:appendix1} for details. 
We parametrized the contour in the mathematical positive sense, i.e.,  the interior of the cell is on the left when going around the contour with increasing $\theta$.
In the numerical implementation, we used the rescaled arc length coordinates, which we denoted again by $\theta$. This gives
\begin{equation}\label{eq:arclength}
    \Vert \partial_\theta \Phi_k(\theta)\Vert \equiv \frac{L_k}{2\pi},\quad \text{with}
    \quad L_k = \int_0^{2\pi} \Vert \partial_\theta \Phi_k(\theta)\Vert \rd\theta,
\end{equation}
denoting the length of contour $\Gamma_k$. Note that $\Phi_k$ is only uniquely determined up to a phase shift, i.e., for every $\Phi_k$ and $\tau$, also $\Phi_{k,\tau}(\cdot) = \Phi_k(\cdot +\tau)$ is a valid parametrization of $\Gamma_k$. The phase shift was chosen by an additional requirement in the next section. 

The smooth parametrization $\Phi_k$ allowed us compute local quantities along the contour, e.g.,  its curvature
\begin{equation}\label{eq:curvature}
\kappa = \frac{R_{\pi/2}\partial_\theta \Phi_k(\theta) \cdot \partial^2_\theta \Phi_k(\theta)}{\Vert \partial_\theta\Phi_k(\theta)\Vert^3},
\end{equation}
where $R_{\pi/2}$ is an anti-clockwise rotation by $\pi/2$. As global quantity, we determined the center of mass $C_k$ of contour $\Phi_k$ as
\begin{equation}\label{eq:centerofmass}
C_k = \frac{1}{2\pi} \int_0^{2\pi} \Phi_k(\theta) d\theta.
\end{equation}
Since the segmentation points were rather densely spaced over the contour,  they well constrained the smooth contours. 
Based on the kernel representation, all geometric quantities,
such as arc length and curvature were defined intrinsically for each
contour, and may be easily computed numerically and with high precision. 

Connecting the contours along the time axis, however, is not intrinsically well defined, and is bound to choices. 
In a first step, we constructed a global coordinate flow, which served as a reference frame
for further local flows to be defined subsequently. In the limit of infinitely densely sampled contours, this global coordinate system results in a mapping 
\begin{equation*}
    \Phi: [0, T]  \times [0,2\pi) \to \mathbb{R}^2,
    \quad (t,\theta) \mapsto \Phi(t,\theta).
\end{equation*}
Vice versa, any coordinate system defines a coordinate flow over the tube of contours, i.e., the cell contours in 2d mapped into a 3d space-time coordinate system. If $p_0=(x_0, y_0)$ is a point on the first contour at $t=0$ and if 
$\theta_0$ is its arc length coordinate,  
then $t\mapsto \Phi(t,\theta_0)$ corresponds the movement of $p_0$ over the space--time tube of contours, see \nameref{s:amoebatube}. 
Of note, this coordinate flow is a theoretical construct that allows us to analyze amoeboid contour dynamics and should not be misinterpreted as a flow of specific membrane lipids or proteins.
Nevertheless, such a global coordinate system is useful and allows for graphical visualization of the contours and any locally defined quantity in form of a kymograph.

\begin{figure}
\begin{center}
\showFigure{
\includegraphics[width=0.8\linewidth]{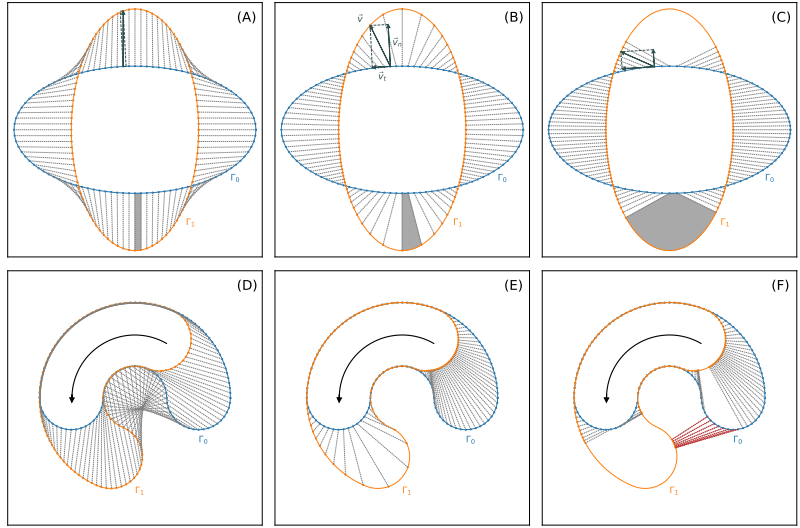}
}
\end{center}
\caption{\textbf{Virtual marker flows} for two test cases w.r.t different degrees of regularization: strongly regularized \textbf{(A, D)}, weakly regularized \textbf{(B, E)}, and non-regularized \textbf{(C, E)}.
In panel (A)--(C), a stretching effect is illustrated as gray area. Furthermore, mapping violations are highlighted as red lines \textbf{(F)}.
}
\label{fig:geoplot}
\end{figure}

%--------------------------------------------------------------------------------------------------------------------------------------------------
\paragraph{Biophysical motivation \& description of our contour dynamics approach}
After obtaining smooth contours at discrete time steps the question of how to link them in time remains open. Mapping a marker on one contour to its nearest point on the consecutive contour is one desirable feature; achieving a regularly spaced distribution of markers on the consecutive contour is another desirable feature. Typically, a regularization parameter is used to balance the two features.
In Fig~\ref{fig:geoplot}, mappings defined by differently regularized flows of virtual markers are shown for two consecutive contours:  
a strong regularization (left column), a weak regularization (middle column), and no regularization at all (right column). While the strongly regularized flow preserves the distances of neighboring virtual markers, the non-regularized flow is defined by the shortest paths from the first contour to the second.

Additionally, in panel (B), the velocity $\vec{v}(\theta)$ that propagates a virtual marker $\theta$ over a particular distance to the next contour, is illustrated as well as its decomposition into an outward-pointing normal velocity $\vec{v}_n(\theta)$ and a tangential velocity $\vec{v}_t(\theta)$.

For the weakly and non-regularized cases, a stretching effect of virtual markers can be observed for expanding parts (gray area), whereas clustering effects of virtual markers occur at contracting regions. Assuming infinitesimally small time steps, the ``stretching rate'' at a virtual marker $\theta$ that arises from transitioning from one contour to the next one is given by
\begin{equation}\label{eq:dilrate}
    d(\theta) = \frac{\partial v_t(\theta)}{\partial\theta} + \kappa(\theta)\cdot v_n(\theta),
\end{equation}
where $v_t$ and $v_n$ denote the (scalar) speed in tangential and normal direction, respectively, and $\kappa$ the curvature.
In the following sections, we show that this kind of ``stretching rate'', in the sequel termed instantaneous dilation rate, is a useful quantity to identify active regions of the contour during cell motility.

While the normal component of the velocity can be approximated easily by taking the normal distances from the first contour to the second divided by $\delta t$, the tangential component remains unknown and therefore requires additional attention. Naive flows, such as the shortest path flow from Fig~\ref{fig:geoplot} (right column) or the flow merely based on the normal component, can produce topological mapping violations. An example of such a mapping violation can be seen in panel (F) where the order of virtual markers on the second contour is inconsistent with the previous one.
\customspace

In general, additional constraints on the tangent component are necessary with the aim of preserving an evenly-spaced distribution of virtual markers over time.
In \cite{machacek_morphodynamic_2006}, these additional constraints were formulated as a mechanistic spring model, providing better computation times than the more commonly used LSM. However, it is mentioned that for strong shape deformations this mechanistic model easily fails because of mapping violations.
This approach, as well as ours, is based on a cost functional minimizing the trajectories of virtual markers to the next contour while penalizing the distances of neighboring virtual markers. While in the mechanistic spring model the penalizing distance is measured in $\R^2$, the space of contours, our approach measures the penalizing distance in $[0, 2\pi)$, the space to parametrize the contour. This key difference resolves the issue of mapping violations during large deformations in our approach.

In \nameref{s:compareregularization}, contour mappings for each of these two regularization schemes are displayed showing the advantageous stability of our developed $\mathcal{S}^1$ regularization during membrane spikes and other large shape deformations.
In \nameref{s:largeshapedeformations}, a selection of contour mappings between two frames with larger shape deformations are displayed. The underlying microscopy data was recorded with an imaging rate of $\delta t\approx 3.13s$.
\customspace

Our method combines and improves different aspects of previous approaches:
\begin{itemize}
    \item Spatio-temporal tracking of a fixed number of virtual markers,
    \item Distances between virtual markers flexibly adjustable in terms of a single regularization,
    \item Faster than computationally expensive approaches such as LSM,
    \item Capability to capture large shape deformations/avoidance of mapping violations,
    \item Simplicity regarding the interpretation \& number of parameters.
\end{itemize}
Another key feature of our approach is the usage of two different flows, separating the underlying coordinate system, defined by one (strongly regularized) flow, and the dynamic quantities of interest, which are defined separately, e.g., by a weakly regularized flow.

%--------------------------------------------------------------------------------------------------------------------------------------------------
\paragraph*{The maximal correlation coordinate system (MCCS).}
The starting point are the parameterized contours $\Phi_k$ in Eq~\eqref{eq:contourmaps} for $k=0,\ldots,K-1$. 
To make the influence of the sampling rate more prominent, we also used the notation 
\begin{equation*}
    \Phi(k\delta t,\theta) = \Phi(t_k, \theta) = \Phi_k(\theta).
\end{equation*}
As stated above, the parametrization of $\Phi_k$ is only determined up to a phase shift by  
Eq~\eqref{eq:arclength}. 
We finally chose the phase shift and therefore the parametrization of $\Gamma_{k+1}$ by minimizing the distance to the previous contour $\Gamma_{k}$ in a mean squared sense, i.e., 
\begin{equation} \label{eq:phaseshift}
   \tau_{k+1} =  \argmin_{ \tau } \int_0^{2\pi} 
    \Vert \Phi_{k+1}(\theta-\tau) - \Phi_{k}(\theta)\Vert^2  \rd \theta.
\end{equation}
We may alternatively interpret Eq~\eqref{eq:phaseshift} as optimizing the cross covariance between the two contours when interpreted as vector-valued functions
\begin{equation*}
\tau_{k+1} =  \argmin_{ \tau } \int_0^{2\pi}\Phi_{k+1}(\theta-\tau) \cdot   \Phi_{k}(\theta) \,\rd\theta.
\end{equation*}
In the sequel, we used $\widetilde\Phi_{k+1}(\cdot) = \Phi_{k+1}(\cdot+\tau_{k+1})$ and omitted the tilde for ease of notation. Effectively, choosing the phase shift amounts to fixing the 'zero point' $\Phi_{k+1}(0)$ on $\Gamma_{k+1}$. 

This is the coordinate system that from now on is used to represent the contour geometry, i.e., each scalar quantity $q$ defined on the contour $\Gamma_k$ with $q=q(t_k,(x,y))$ and $(x,y)\in\Gamma_k$ as function $(k,\theta) \mapsto q(k\delta t,\theta)$ of discrete time and continuous space can be represented w.r.t.\ the chosen coordinate system $\Phi$.

%--------------------------------------------------------------------------------------------------------------------------------------------------
\paragraph*{The Eulerian and Lagrangian points of view.}
Any flow that maps the  contour $\Gamma_k$ into $\Gamma_{k+1}$ is 
determined by a mapping which describes the translation along the contour $\phi_k$. 
To ensure that 
$\theta \mapsto \phi_k(\theta)$ is a one-to-one map, we required in addition
\begin{equation} \label{eq:donotallowformappingviolations}
    \partial_\theta \phi_k(\theta) > 0.
\end{equation}
An example of a mapping violation, i.e. a position $\theta$ with $\partial_\theta \phi_k(\theta) < 0$, is illustrated in Fig~\ref{fig:geoplot}F, where the order of virtual markers on the following contour is inconsistent with the previous one.
The iteration
\begin{equation*}
    \theta_{k+1} =  \phi_k(\theta_k)
\end{equation*}
describes the trajectory $(\theta_k)_{k=0,\ldots,K-1}$ of the starting point at coordinate $\theta_0$ on the first contour 
in our coordinate system.  
This approach to visualize the flow shall be called the Eulerian point of view, since it describes the translation vector
field from $\Gamma_k$ to $\Gamma_{k+1}$ in the coordinate system of $\Gamma_k$: 
\begin{equation} \label{eq:V_k}
    \Phi_{k+1}(\phi_k(\theta)) =  \Phi_k(\theta) + \delta t V_k(\theta).
\end{equation}
The Lagrangian point of view instead describes the flow in the coordinate system it generates, which is different from our MCCS. 
Denote the coordinate of a point on the first contour by its angle coordinate $\theta_0$, and let $\chi_k$ be the mapping of $\Gamma_0$ to $\Gamma_k$ recursively defined by 
\begin{equation*}
\chi_{k+1}(\theta_0)  = \phi_k( \chi_k(\theta_0)), \quad \chi_0(\theta_0)=\theta_0.
\end{equation*}
This gives $\chi_k(\theta_0)=\theta_k$ for $k=0,\ldots,K-1$. The Lagrangian description $\Xi(t_k, \theta_0)\in \mathbb{R}^2$ 
is linked to the Eulerian description via 
\begin{equation*}
    \Xi( t_k,\theta_0) = \Phi(t_k, \chi_k(\theta_0)). 
\end{equation*}
Both points of view are useful to understand and describe a flow over the contour. 
The translation vector $W_k$ in Lagrangian coordinates is simply
\begin{equation*}
    \Xi(t_{k+1},\theta_0 ) = \Xi(t_k,\theta_0) +  \delta t W_k(\theta_0)
\end{equation*}
and is linked to the Eulerian description via
\begin{equation*}
    V_k(\chi_k(\theta_0)) = W_k(\theta_0).
\end{equation*}
We used the functions $\phi_k$, $V_k$ and $W_k$ interchangeably to specify
the flow from $\Gamma_k$ to $\Gamma_{k+1}$.

%--------------------------------------------------------------------------------------------------------------------------------------------------
\paragraph{Transport along the flow.}
%--------------------------------------------------------------------------------------------------------------------------------------------------
For any flow, we may define the instantaneous 
dilation rate \LD\ of the flow. 
Considering two infinitesimally nearby points on contour $\Gamma_k$, we see that the local relative 
dilation/contraction factor is obtained from $\phi_k$ via
\begin{equation}\label{eq:dilationrate}
    \LD_k(\theta) \delta t=\log( \partial_\theta \phi_k(\theta) ).
\end{equation}
Note that our global coordinate system MCCS induces a flow with uniform dilation rate. 
To describe the transport of a density under the flow, consider points on the contour $\Gamma_k$ 
that are distributed according to a density $\mu_k(\theta)\rd\theta$. 
Under the flow induced by $\phi_k$ this density 
changes according to 
\begin{equation}\label{eq:distmapping}
    \mu_{k+1}(\phi_k(\theta)) \rd\theta = 
    \frac{\mu_k(\theta)\rd\theta}{\partial_\theta\phi_k(\theta)}
\end{equation}
Starting from $\mu_0(\cdot) \equiv 1/(2\pi)$, this defines the transported density on all contours.
In the Lagrangian picture this transport preserves the density $\mu_0(\cdot)$, which follows from the fact that by definition the starting angle does not change under the flow. The density $\mu_k$ can be written in Lagrangian coordinates as
\begin{equation} \label{eq:lagrangedens}
    \mu_k(\chi_k(\theta_0)) = \frac{1}{2\pi\cdot\partial_{\theta_0} \chi_k(\theta_0)}.
\end{equation}

%--------------------------------------------------------------------------------------------------------------------------------------------------

\paragraph{A regularizing family of flows.}
%--------------------------------------------------------------------------------------------------------------------------------------------------
We next defined a family of \textit{local} mappings $\phi_k$ between successive contours that 
yields a compromise between the shortest path flow
and the uniform dilation coordinate flow, presented in the two end-member cases in Fig~\ref{fig:geoplot}. Another suitable name for shortest path flow is reversed normal flow since the incoming trajectories under this flow are always orthogonal to the successive contour.

The mean squared velocity of the flow (with respect to a density $\mu_k$) is given as
\begin{equation} \label{eq:F_k_phi}
    F_k[\phi_k] = \int_0^{2\pi} 
    \left\Vert \frac{\Phi_{k+1}(\phi_k(\theta) ) - \Phi_k(\theta)}{ \delta t } \right\Vert^2 \mu_k(\theta)\rd\theta = \int_0^{2\pi}\Vert V_k(\theta)\Vert^2 \mu_k(\theta) \rd\theta .
\end{equation}
The normal flow from contour $\Gamma_k$ to $\Gamma_{k+1}$ is the flow that departs from the first contour in the normal direction until it intersects with the second contour. The normal flow from $\Gamma_{k+1}$ to $\Gamma_k$ shall be called the reversed normal flow from $\Gamma_k$ to $\Gamma_{k+1}$. 
This is the unconstrained minimizer of $F_k$. If there are no intersections 
of flow lines, it defines a one-to-one mapping between the two contours.
In general, however, direct minimization of the functional $F_k$ does not yield a valid flow because of self-intersections, and as a consequence, the induced map is multiple-valued. We therefore need to regularize the flow. A natural requirement is that the flow tends to enforce non-uniformly distributed points on contour $\Gamma_k$ towards more uniformly distributed points on contour $\Gamma_{k+1}$. We proposed to quantify the 
degree of non-uniformity of a distribution $\mu(\theta) \rd\theta$ by means of 
\begin{equation}\label{eq:nonuniformitymeasure}
    U[\mu] = \int_0^{2\pi} \frac{\rd\theta}{\mu(\theta)}.
\end{equation}
Since any distribution satisfies $\int_0^{2\pi} \mu(\theta)=1$ and $\mu(\theta)>0$, the minimizer actually corresponds to the uniform distribution. Other measures of non-uniformity are possible, for instance
the entropy 
\begin{equation*}
    S[\mu] = \int_0^{2\pi} \mu(\theta) \log( \mu(\theta)) d\theta.
\end{equation*}
This kind of regularization has been proposed in \cite{otto_2001} by Otto, where the optimal flow is understood as a gradient flow of the entropy potential with respect to the Wasserstein transport distance of the marker density.
Also very appealing, in this paper, we used the characterization in Eq~\eqref{eq:nonuniformitymeasure}, since it leads to a more readily tractable numeric quadratic minimization problem.
In terms of the defining mapping $\phi_k$, the functional $U$ reads 
\begin{equation}
    U_k[\phi_k] = \int_0^{2\pi} \frac{\partial_\theta \phi_k(\theta)^2}{\mu_k(\theta)}\rd\theta,
\end{equation}
as follows from Eq~\eqref{eq:distmapping}.
The regularized flow is defined as the flow that minimizes a compromise between both cost functions
\begin{equation}\label{eq:regularization}
    \phi_{k,\lambda} = \argmin_{\phi_k} \; F_k[\phi_k] + \lambda  U_k[\phi_k], \quad \lambda>0.
\end{equation}
Note that $H_k[\phi_k]=F_k[\phi_k] + \lambda  U_k[\phi_k]$ depends on both, $\phi_k$ and the measure $\mu_k$. 
When iterating over all contours, one needs to update the measure before optimizing the flow for the next time step. There are two end-member cases for the regularized flow:
\begin{itemize}
\item For large $\lambda$ the optimal flow essentially immediately uniformizes the density of the initial contour. Thereafter, it is the uniform stretching flow that minimizes the mean square distances between the contours. 
This is precisely the coordinate flow defined before. 
\item For small $\lambda$ the optimal flow allows for arbitrary local stretching rates to minimize the point-wise distance.
Here the limit defines the regularized shortest path flow, respectively, the regularized reverse normal flow.
\end{itemize}
Note that straightforward pointwise minimization of the flow distance 
from $\Gamma_k$ to 
$\Gamma_{k+1}$ may lead to overlapping connections and hence singular mappings between the contours. 
If instead, we regularize with small $\lambda$, such overlaps are avoided.

%--------------------------------------------------------------------------------------------------------------------------------------------------
\paragraph{The virtual marker picture.}
%--------------------------------------------------------------------------------------------------------------------------------------------------

For numerical implementation, we discretized the cost functionals using the concept of virtual markers on the contours.
The virtual markers are a discretized version of the Lagrangian coordinates.
Since  $\mu_k$ is the transported density, 
the first cost functional for $\phi_k$  in the Lagrangian point of view using Eq~\eqref{eq:lagrangedens} is given by 
\begin{align*}
    F_k[\phi_k]
    &=\frac{1}{2\pi}\int_0^{2\pi}\Vert V_k(\chi_k(\theta_0))\Vert^2 \rd\theta_0\\
    &=\frac{1}{2\pi\delta t^2}\int_0^{2\pi}\left\Vert \Phi_{k+1}(\phi_k(\chi_k(\theta_0)) ) - \Phi_k(\chi_k(\theta_0)) \right\Vert^2 \rd\theta_0,
\end{align*}
while the second functional using Eq~\eqref{eq:lagrangedens} is given by 
\begin{equation} \label{eq:U_k}
    U_k[\phi_k]
    = 2\pi\int_0^{2\pi} |\partial_{\theta_0}\chi_{k+1}(\theta_0)|^2 \rd\theta_0.
\end{equation}
See \nameref{s:appendix1} for details of the derivation. Both equations are well suited for a discrete numerical approximation for a given function $f$ and initially equidistant $\theta_{0,i} = 2\pi i/N$ with $i=0,\ldots,N-1$ using 
\begin{equation*}
    \sum_{i=0}^{N-1}  f(\theta_{0,i}) (\theta_{0,i+1}-\theta_{0,i} ) \simeq \int_0^{2\pi} f(\theta_0) \rd\theta_0.
\end{equation*}
If we now approximate the continuous mapping $\phi_k$ by its 
discrete values on the virtual marker points $\theta_{k} = (\theta_{k,0},\ldots,\theta_{k,N-1})$ with
\begin{equation}  \label{eq:iterativevirtualmarkers}
\theta_{k+1,i} = \phi_k(\theta_{k,i}),
\end{equation} 
then the first cost function may be approximated as
\begin{equation}\label{eq:first_term}
    F_k[\phi_k] \simeq F_k\big[\theta_{k+1}|\theta_{k}\big]
    = \frac{1}{N \delta t^2} \sum_{i=0}^{N-1}
    \big\Vert\Phi_{k+1}(\theta_{k+1,i}) -\Phi_{k}(\theta_{k,i})\big\Vert^2
\end{equation}
and the second cost function as
\begin{equation}\label{eq:second_term}
    U_k[\phi_k] \simeq U_k\big[\theta_{k+1}| \theta_{k}\big]
    = N \sum_{i=0}^{N-1} \big\vert\theta_{k+1,i+1}-\theta_{k+1,i}\big\vert^2.
\end{equation}
For the entropy based measure of uniformity, consider a collection of points $\theta_{k,0},\ldots,\theta_{k,N-1}\sim \mu_k$ on $\Gamma_k$ that are distributed according to the density $\mu_k(\cdot)$ (not necessarily uniform). 
Then for any function $f$, it is
\begin{equation*}
    \frac{1}{N}\sum_{i=0}^{N-1}  f(\theta_{k,i}) \simeq \int_0^{2\pi} f(\theta) \mu_k(\theta) \rd\theta,
\end{equation*}
yielding
\begin{equation*}
    S_k[\phi_k] \simeq S_k\big[\theta_{k+1}| \theta_{k}\big] = \frac{1}{N} \sum_{i=0}^{N-1} \log( \theta_{k+1,i+1}-\theta_{k+1,i}).
\end{equation*}
In this discrete virtual marker approximation, the local dilation rate at $\theta_{i,k}$, also called the local dispersion,  reads
\begin{equation} \label{eq:localdispersion}
    \LD_{k,i} = \frac{1}{\delta t}\log \frac{\theta_{k+1,i+1} - \theta_{k+1,i}}{\theta_{k,i+1} - \theta_{k,i}}.
\end{equation}
Please note that for infinitesimally small time steps another formulation of this rate is given by Eq~\eqref{eq:dilrate}. Especially for level set methods, this formula is more applicable.
Finally, the discrete optimization problem is given by
\begin{equation}\label{eq:discreteregularization}
    \phi_{k,\lambda} = \argmin_{\phi_k} \; F_k\big[\phi_k(\theta_{k})| \theta_{k}\big] + \lambda  U_k\big[\phi_k(\theta_{k})| \theta_{k}\big], \quad \lambda>0.
\end{equation}
Note that in the discrete optimization problem we do not pose any requirements on the space of transformations $\phi_k$ ensuring condition Eq~\eqref{eq:donotallowformappingviolations}. If $\phi_k(\theta_{k,i+1})-\phi_k(\theta_{k,i})$ $= \theta_{k+1,i+1}-\theta_{k+1,i} \leq 0$ for some $k$, we say that $\phi_k$ exhibits mapping violations. 
%

%--------------------------------------------------------------------------------------------------------------------------------------------------
\paragraph{Marker re-initialization for weakly-regularized contour flows.}
%--------------------------------------------------------------------------------------------------------------------------------------------------

In general, the distribution of virtual markers $\theta_k\sim\mu_k$ on $\Gamma_k$ depends on the initial distribution of virtual markers $\theta_0\sim\mu_0$ on $\Gamma_0$, since the density  
$\mu_k$ results from the transport of $\mu_0$ by the flow. As a consequence, functions derived from the local contour flow like, e.g., the local dilation rate along $\Gamma_k$, may depend on the initial distribution on $\Gamma_0$. By re-initializing the distribution of virtual markers on $\Gamma_k$ for any $k=1,\ldots,K-1$, this dependence may be removed. A natural choice is to re-initialize with the uniform distribution, i.e., using 
\begin{equation}  \label{eq:reinitializedvirtualmarkers}
\theta_{k+1,i} = \phi_k(\xi_i)
\end{equation} 
with $\xi_i=2\pi i$ for $i=0,\ldots,N-1$ instead of Eq~\eqref{eq:iterativevirtualmarkers}. To approximate the local flow $\phi=(\phi_k)_{k=0,\ldots,K-1}$ between contours, we thus solved the re-initialized optimization problem 
\begin{equation}\label{eq:reinitregularization}
    \phi_{k,\lambda} = \argmin_{\phi_k} \; F_k\big[\phi_k(\xi)| \xi\big] + \lambda  U_k\big[\phi_k(\xi)| \xi\big], \quad \lambda>0.
\end{equation}
with $\xi=(\xi_i)_{i=0,\ldots,N-1}$. For the local dispersion, e.g., this resulted in
\begin{equation} \label{eq:reinitializedlocaldispersion}
    \LD_{k,i} = \frac{1}{\delta t} \log \frac{\phi_{k,\lambda}(\xi_{i+1})-\phi_{k,\lambda}(\xi_i)}{\xi_{i+1} - \xi_i}
\end{equation}
with no dependence on the initial distribution of markers on $\Gamma_0$. 

%--------------------------------------------------------------------------------------------------------------------------------------------------
\paragraph{Algorithmic workflow.}
%--------------------------------------------------------------------------------------------------------------------------------------------------

We summarize the proposed numerical workflow:
\begin{enumerate}
\item Given the segmented contours $\gamma_0,\dots,\gamma_{K-1}$ as in Eq~\eqref{eq:data}, determine the continuous representations $\Phi_k$ defined in Eq~\eqref{eq:contourmaps} satisfying the conditions Eq~\eqref{eq:arclength}.
\item Determine contour based quantities like the curvature in Eq~\eqref{eq:curvature} or the center of mass in Eq~\eqref{eq:centerofmass}.
\item To determine the (strongly-regularized) coordinate flow, consider $N$ equally spaced markers 
\begin{displaymath}
\theta_0 = \left( \theta_{0,i}\right)_{i=0,\ldots,N-1}\qquad \theta_{0,i} = \xi_i = \frac{2\pi i}{N}
\end{displaymath}
on the initial contour and choose a large $\lambda$ value. Iteratively solve the regularization problem in Eq~\eqref{eq:reinitregularization} to determine the coordinate markers $\theta_{k+1}$ for $\Gamma_{k+1}$ from the coordinate markers $\theta_k$ for $\Gamma_k$. Since both $\theta_{k+1}$ and $\theta_k$ are approximately equally spaces, solving the minimization problem amount to choosing $\theta_{k+1,0}$. 

\item To determine the (weakly regularized) local flow $\phi=(\phi_k)$ between successive contours, choose a small $\lambda$ value and solve the regularization problems in Eq~\eqref{eq:reinitregularization} to determine the coordinates $\theta_{k+1,i}=\phi_{k,\lambda}(\xi_i)$ on $\Gamma_{k+1}$ based on $N$ equally spaced markers $\xi_0,\ldots,\xi_{N-1}$ on $\Gamma_k$. 
\item Determine contour flow based quantities like the local dispersion in Eq~\eqref{eq:reinitializedlocaldispersion} or the local motion:
\begin{equation}
    \LM_{k,i} =
    \Vert V_k(\xi_i) \Vert
    =\frac{\Vert\Phi_{k+1}(\phi_{k,\lambda}(\xi_i))-\Phi_k(\xi_i)\Vert}{\delta t},
\end{equation}
see also Eq~\eqref{eq:V_k}.
\item Map all quantities, based on contour flow as well as contour only, onto the (strongly regularized) global flow.
\end{enumerate}

\paragraph{Numerical implementation and reproducibility.}

All methods presented in this article are fully accessible as an open source Python-based toolbox \texttt{AmoePy} \cite{amoepy_2020}. 
Implementations of the above methods and additional routines necessary to reproduce the figures and results of this article are part of this toolbox. Furthermore, \texttt{AmoePy} contains:
\begin{itemize}
    \item An object-oriented analysis tool to handle cell contours, i.e., to shorten, extend, or manipulate existing data sets and to extract additional geometric quantities (e.g. area, perimeter, and normal vectors along the cell contour),
    \item A python class to perform Gaussian process regressions in 2d (space) and 3d (space-time) with different selectable kernel functions,
    \item Multiple testing routines based on artificial test data and experimental data,
    see \nameref{s:appendix1} and \nameref{s:testcases} for more details,
    \item Several routines generating videos of cell tracks with corresponding kymographs and occurring expansion/contraction patterns,
    \item A detailed documentation generated by the Python documentation tool Sphinx,
    \item A graphical user interface able to compute and present kymographs
\end{itemize}
\texttt{AmoePy}, as well as its graphical user interface, is updated regularly. Future outcomes of our ongoing research regarding for example cell segmentation routines and a forward model to simulate amoeboid cell motility will be also added to \texttt{AmoePy}.
\customspace

\begin{figure}
\begin{center}
\showFigure{
\includegraphics[width=0.85\linewidth]{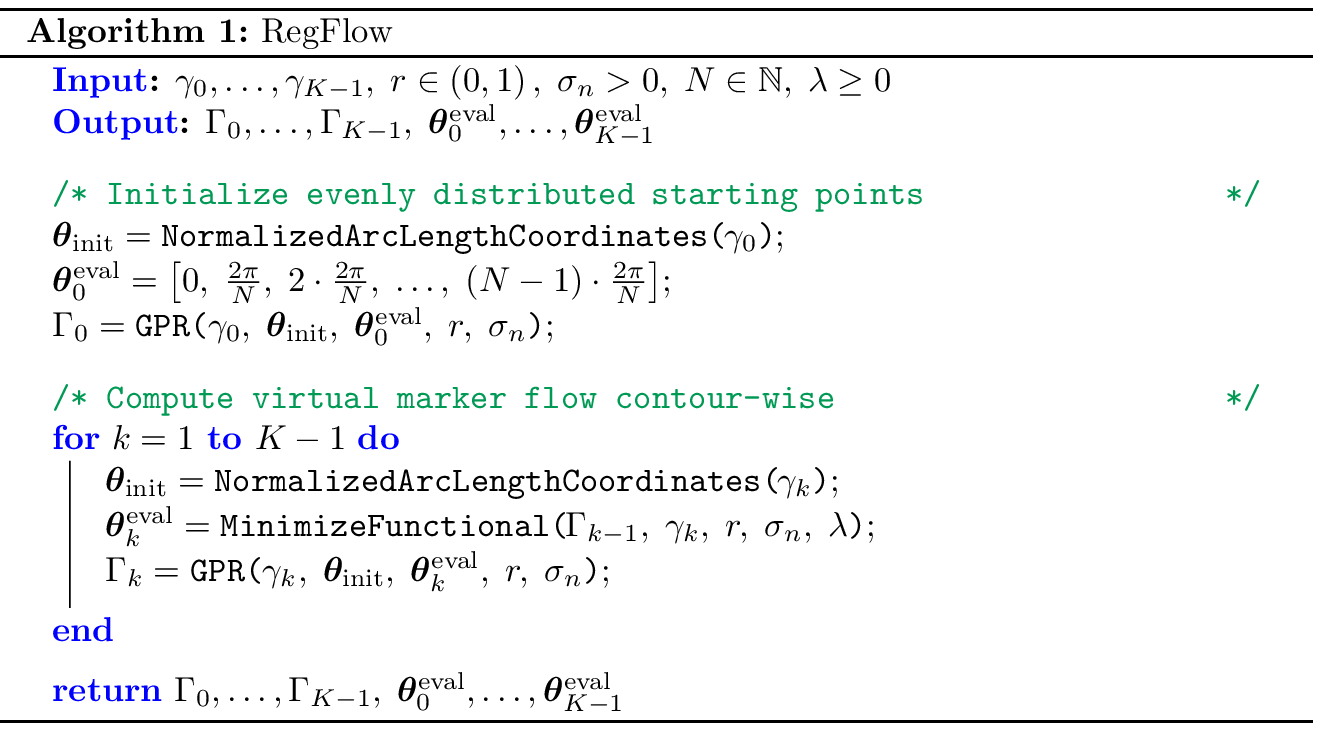}
}
\end{center}
\caption{\textbf{Algorithm to compute regularized flows.} The algorithm input consists of three parameters $r$, $\sigma_n$, and $N$ regarding the Gaussian process regression, a regularization parameter $\lambda$ and the initial (segmented) contours $\gamma_0,\dots,\gamma_{K-1}$. The regularized flow is described by the output variables $\Gamma_0,\dots,\Gamma_{K-1}$ denoting smooth contours evaluated at a finite number of coordinate markers $\boldsymbol{\theta}_0^\text{eval},\dots,\boldsymbol{\theta}_{K-1}^\text{eval}$.}
\label{fig:pseudocodeRF}
\end{figure}

\begin{figure}
\begin{center}
\showFigure{
\includegraphics[width=0.85\linewidth]{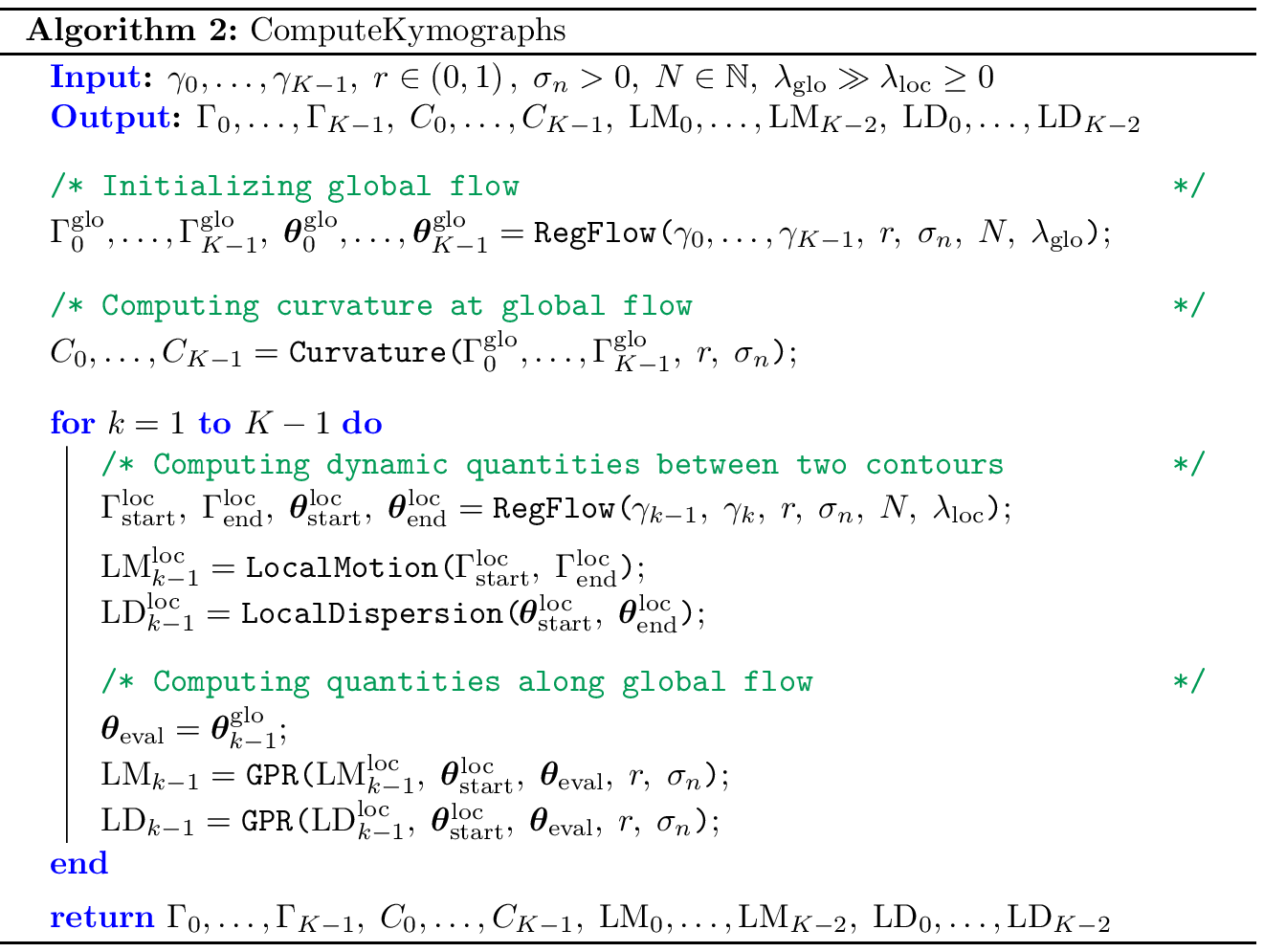}
}
\end{center}
\caption{\textbf{Algorithm to compute curvature, local motion and local dispersion.} The two regularization parameters $\lambda_\text{glo}$ and $\lambda_\text{loc}$ are used in the RegFlow routine (see Fig~\ref{fig:pseudocodeRF}) to determine the global (strongly regularized) and local (weakly regularized) flow. The output comprises the smoothed contours $\Gamma_0,\dots,\Gamma_{K-1}$ as well as geometric quantities of interest such as curvature, local motion, and local dispersion.}
\label{fig:pseudocodeCK}
\end{figure}

The algorithm starts with initializing equally spaced markers, representing the starting points of the flow (see also the algorithmic workflow). Subsequently, the optimization problem in Eq~\eqref{eq:reinitregularization} is solved contour-wise by gradient descent.
Here, it is beneficial that the GPR also provides these gradients as a by-product, which decreases the computation time drastically; see \nameref{s:appendix1} and \nameref{s:pseudocodeMF} for more details. Fig~\ref{fig:pseudocodeRF} shows the pseudocode to compute a regularized flow for a given regularization parameter $\lambda$. Being able to determine the regularized flow for a given parameter $\lambda$ is the prerequisite to finally compute the global (strongly regularized) and local (weakly regularized) flows. These are based on the two regularization parameters $\lambda_\text{glo}\gg\lambda_\text{loc}>0$. The local quantities are then represented in the coordinate system of the global flow, yielding the kymograph representation. A pseudocode describing the computation of these kymographs is shown in Fig~\ref{fig:pseudocodeCK}.

\customspace

In contrast to level set and electrostatic methods, the optimization approach presented here is not relying on the computation of intermediate field lines or contours for sufficiently small grid sizes. Notably, level set methods are computationally more expensive because of the time integration of virtual markers along these field lines\cite{machacek_morphodynamic_2006,tyson_high_2010}.
On the other hand, our algorithm contains additional steps such as the re-initialization of virtual markers via Gaussian process regression and the computation of kymograph quantities which may lead to slower computation times than other empirical mapping algorithms. A direct comparison of computation times of these methods is rather difficult since some algorithms rely on a varying number of virtual markers or a more image-based implementation (e.g. in \texttt{ImageJ}). Moreover, the computation time depends on varying configurations of these methods, e.g., the resolution of the underlying grid in LSM, or the number of iterations/tolerance parameters used in optimization approaches such as ours.
Unfortunately, a ``biological true'' mapping does not exist, with which the accuracy of each of those methods can be measured.

Nevertheless, we provide the computation times of our algorithm ``RegFlow'' shown in Fig~\ref{fig:pseudocodeRF} on a standard computer, see \nameref{s:comptime}. For a cell track of 500 contours and 400 virtual markers, the computation time for generating the mapping between two successive contours was measured for different regularization parameters $\lambda$. Not surprisingly, the computation time decreases for the end-member cases $\lambda\to 0$ and $\lambda\gg 0$, where the underlying functional is only defined by one leading term. 
For each choice of $\lambda$, the median of the computation time is below $2s$ where most of the cases required sub-second computation times. Furthermore, no mapping violations occurred during the cell track despite larger shape deformations and membrane spikes. Therefore, our method has the potential to be used specifically for cell motility models that rely on a fast computation of stable virtual marker trajectories.

% ===================================================================================
\section*{Results}\label{sec:Results}
% ===================================================================================
% --------------------------------------------------------------------------------------------------------------------------------------------------
\subsection*{Degree of regularisation controls distribution of virtual markers}\label{ssec:Regularisation}
% --------------------------------------------------------------------------------------------------------------------------------------------------

\begin{figure}
\begin{center}
\showFigure{
\includegraphics[width=0.7\linewidth]{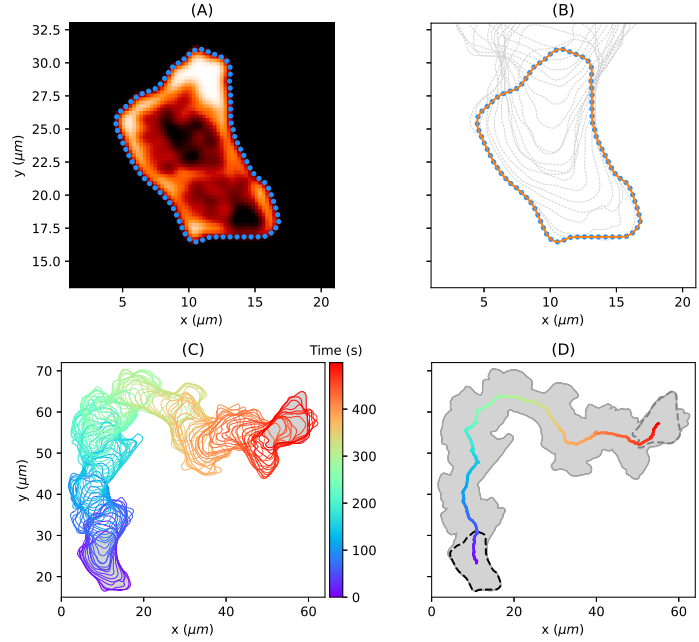}
}
\end{center}
\caption{\textbf{Cell trajectory of a persistently moving amoebae.} \textbf{(A)}~Fluorescence image with closed string of $M=400$ equidistant nodes resulting from the segmentation process; shown is only every fifth segmentation point (blue points).
\textbf{(B)}~Smooth representation $\Phi_k$ of the cell contour (orange line) obtained by spatial Gaussian regression on the segmentation points. Every fifth cell contour is displayed as dashed gray line.
\textbf{(C)}~Entire cell track of $K=500$ cell contours (only every fifth shown).
\textbf{(D)}~Global trace of the cell track (gray area) and the trajectory of the center of mass of the contour (solid line, color coded as in panel (C)). The initial contour is shown as dashed black line and the final contour as dashed gray line.
}
\label{fig:cell-track-009}
\end{figure}

We applied our approach to time-lapse microscopy data of the social amoebae {\it D. discoideum}. Fig~\ref{fig:cell-track-009} illustrates the data acquisition process. For each time point $t_k$, we obtained a sequence of segmentation points  from a fluorescence image (see panel (A) and Eq~\eqref{eq:data}) 
and their corresponding continuous representation $\Phi_k$ (see panel (B) and Eq~\eqref{eq:contourmaps}).
The entire sequence of continuous contours is shown in panel (C), while the trace of the cell track and the center of mass trajectory are shown in panel (D).

In \nameref{s:contourestimate}, estimations of continuous representations for different hyperparameters are shown. In contrast to Fig~\ref{fig:cell-track-009}, the underlying snake of segmented points was obtained from noisy fluorescence image data resulting in many abrupt changes of the contour's direction.
In this context, the curvature for different contour estimates is displayed, highlighting the effect of underfitting and overfitting.
Notably, the GPR provides an automated way of estimating the hyperparameters balancing the model's complexity and its error residuals, see \nameref{s:appendix1} for more information.
This way, an accurate approximation of the cell contour can be obtained even for noisy data while preserving the main characteristics of its shape.

In \nameref{s:temporalresolution}, the effect of different imaging frequencies on the resulting kymographs is shown. In this context, local motion kymographs are computed by using (i) the entire microscopy data (one image/contour per second) and (ii) down-sampled data sets based on every 2nd, 3rd, 5th, and 10th image/contour.
While global characteristics of the cell track are captured even for a lower temporal resolution ($\delta t > 3$), the identification of local membrane changes becomes impossible. We therefore recorded each cell track with an imaging rate of one frame per second ($\delta t=1$). See also \nameref{s:largeshapedeformations}, in which contour mappings for larger shape deformations are displayed. Here, the underlying cell track was recorded with an imaging rate of $\delta t\approx 3.13s$.
\customspace

\begin{figure}
\begin{adjustwidth}{-2in}{0in}
\begin{center}
\showFigure{
\includegraphics[width=0.8\linewidth]{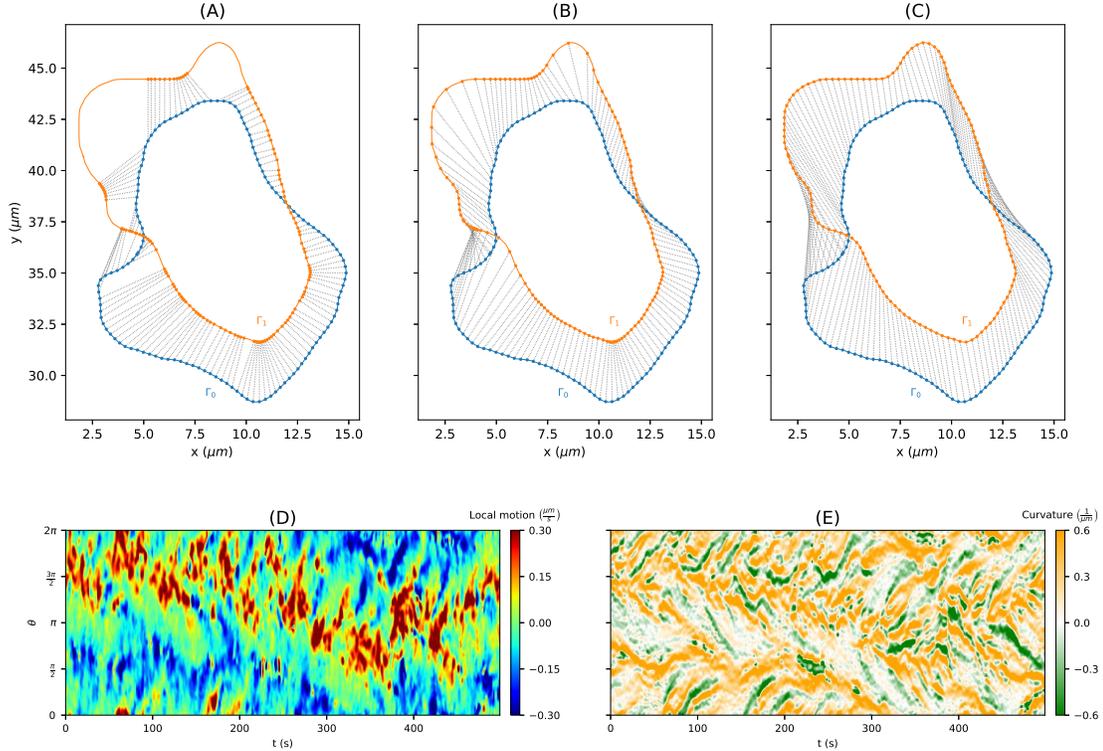}
}
\end{center}
\caption{\textbf{Impact of regularization on the distribution of virtual markers} for \textbf{(A)} no regularization ($\lambda=0$),  \textbf{(B)} weak regularisation with $\lambda=0.1$, and \textbf{(C)} strong regularisation with $\lambda=1000$, illustrated on two frames (roughly $20$s apart for illustration purpose).
Using the strongly regularised so-called coordinate markers as a means to map local characteristics into a kymograph, the lower panel shows the local motion \textbf{(D)} and curvature\textbf{(E)}. The local motion is defined by the magnitude of each mapping vector, which are determined based on the weakly regularised marker flow. All panels correspond to the persistently motile cell of Fig~\ref{fig:cell-track-009}.
}
\label{fig:markerdispersion}
\end{adjustwidth}
\end{figure}

The continuous representations $\Phi_0,\ldots,\Phi_{K-1}$ of all contours are the input to the optimization problem Eq~\eqref{eq:regularization} to determine the regularized contour flow. Fig~\ref{fig:markerdispersion} shows the impact of the regularization parameter $\lambda$ on the virtual marker trajectories (shown for two illustrative contours). In the absence of any regularization (panel (A), $\lambda=0$), virtual markers are thinned out in some regions (linked to expanding areas), while they are clustered in others (in particular at the back of the cell). The case $\lambda=0$ corresponds to minimizing the translation from the first contour to the next, i.e., each virtual marker on the first contour is linked to its nearest neighbor on the second contour. This may result in mapping violations $\theta_{k+1,i+1}-\theta_{k+1,i} \leq 0$ on the second contour.
An example of mapping violations can be seen on the left-hand side in panel (A).
As mentioned earlier, the flow obtained with $\lambda=0$ is defined by the shortest path flow or equally the reversed normal flow.
Thus, instead of taking the trajectories to the nearest neighbors on the next contour, one could also choose the shortest normal vectors from the second contour to the first one.

In the weakly regularized case, virtual marker thinning and clustering is still prominent (see panel (B) with $\lambda=0.1$), but to a lesser extent. Moreover, in the presence of regularization, virtual marker trajectories are interdependent, which results in trajectories without mapping violations. In the limit of strong regularization, the marker points remain uniformly distributed on every contour, while minimizing the overall distance between contours (see panel (C) with $\lambda=1000$).
This makes the strongly regularized virtual marker trajectories an ideal candidate for a time-evolving reference frame and corresponds to the previously defined MCCS coordinate system.

By choosing a strong regularization for the coordinate system, however, information on local contour changes is largely lost. Therefore, we used the strongly regularized case only to determine the coordinate system (set of $N=400$ virtual marker trajectories), while we determined local contour characteristics (e.g. local motion or local dispersion) based on a re-initialized weakly regularized flow. The local characteristics were subsequently represented in the coordinate system obtained from a strongly regularized flow.
In panel (D), the local motion obtained from the re-initialized weakly regularized flow with $\lambda=0.1$ is shown for the same cell track as in Fig~\ref{fig:cell-track-009} and the entire time-lapse microscopy recording (500 frames, $\delta t = 1$s).
The local motion clearly shows regions of fast-moving membrane parts at the leading edge (red areas) and at the back of the cell (blue areas). The curvature kymograph in panel (E) shows characteristic lines of strongly convex (orange) and concave (green) membrane parts.
Inclined lines of curvature (and local motion) may result from adherent parts of the cell moving along the cell contour as well as shifting effects of virtual markers due to arc length changes. It is important to notice that a kymograph depends on the underlying time-evolving coordinate system.

In contrast to the procedure mentioned above, one may also compute local characteristics along the global flow without re-initializing from a uniform distribution of markers after every time step.
However, this is not recommended, because either local information gets lost (strongly regularized flow) or clustering and thinning effects of markers become too prominent (weakly regularized flow);
see \nameref{s:globalflow} for global flows based on different regularization parameters and the resulting kymographs.

As shown in \nameref{s:globalflowtestcase}, we further challenged our algorithm by computing a strongly regularized (global) flow of the cell track from Fig~\ref{fig:cell-track-009} based on a few frames only (8 out of 500). Even under such extreme conditions, the algorithm produced stable virtual marker trajectories, i.e., trajectories without mapping violations.

% --------------------------------------------------------------------------------------------------------------------------------------------------
\subsection*{Kymographs of local properties show characteristic patterns of amoeboid motility}\label{ssec:Comparison}
% --------------------------------------------------------------------------------------------------------------------------------------------------

\begin{figure}
\begin{adjustwidth}{-2in}{0in}
\begin{center}
\showFigure{
\includegraphics[width=0.9\linewidth]{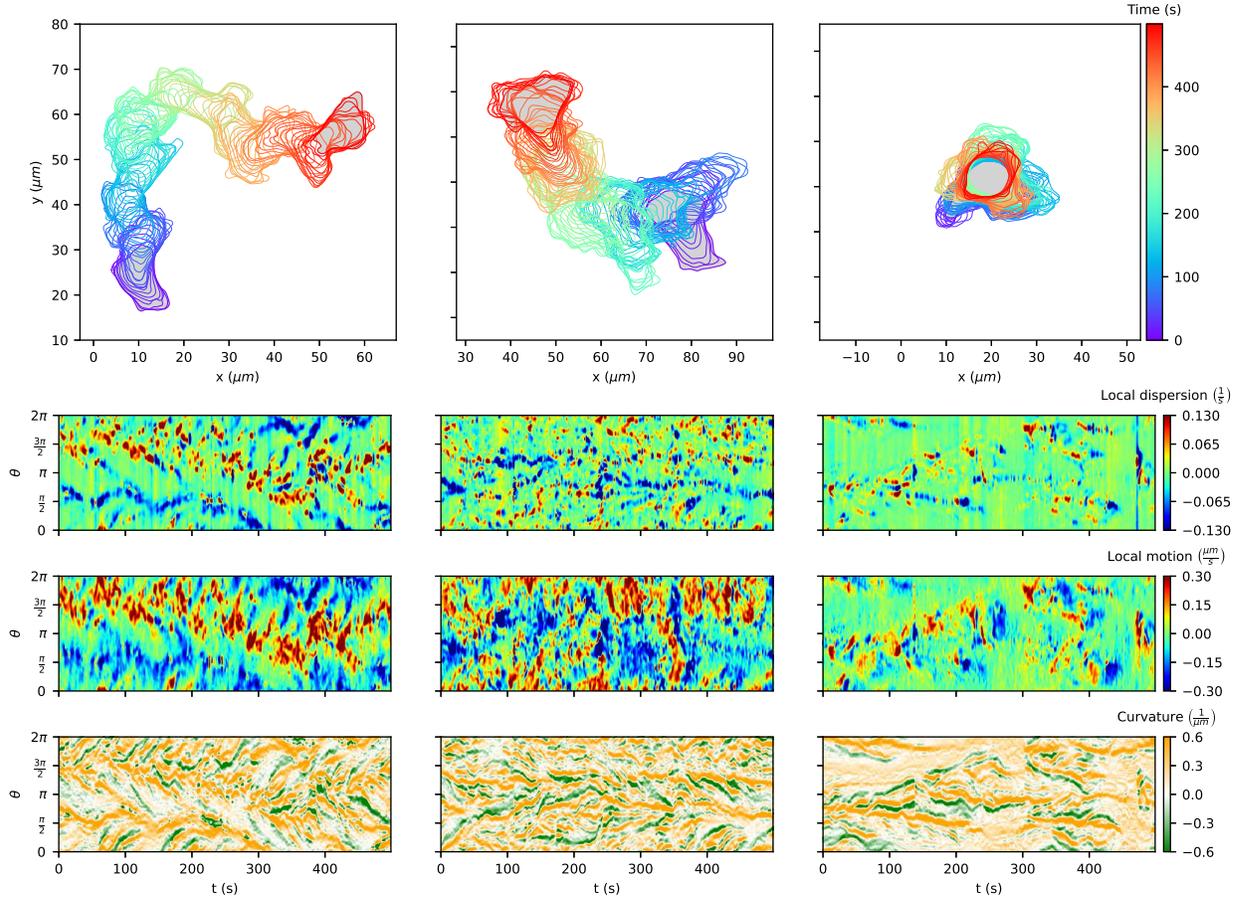}
}
\end{center}
\caption{\textbf{Comparison of different cell tracks of \textit{Dictyostelium discoideum}:} persistently motile (left), weakly motile (middle) and almost stationary (right).
The corresponding kymographs contain information on the local dispersion (left), local motion (middle) and curvature (right). For details see text. 
}
\label{fig:compare_kymo}
\end{adjustwidth}
\end{figure}

The kymographs of local dispersion (\VMD), local motion, and curvature can be used to visualize, analyze, and quantify the expanding activity of a cell track. Note that the three quantities are closely related. Fig~\ref{fig:compare_kymo} shows cell tracks and corresponding kymographs for three different motility patterns: (A) persistently motile cell; (B) weakly motile cell; and (C) an almost stationary cell.
All kymographs are smoothed by a Gaussian filter with a standard deviation of three markers in space and a standard deviation of one contour in time. See \nameref{s:compkymo} for the same kymographs but without smoothing. Moreover, videos of all three cell tracks and their corresponding kymographs can be found in \hyperref[s:009-vid]{S1}--\hyperref[s:018-vid]{S3}~Videos.

For the persistently motile cell, the \VMD~kymograph shows strong (positive) activity in a band-like structure along roughly half of the cell contour (width $\pi$), while the activity of local dispersion is less localized for the weakly motile cell and much less pronounced for the stationary cell. A similar scenario is seen in the local motion kymographs of the three cells. In broad terms, the local motion kymographs show more activity, e.g., areas of red and dark red color, than the \VMD~kymographs. This is also apparent from a correlation plot of the two quantities (see \nameref{s:correlation}).

In the following section, we chose the local dispersion as a basis to identify expansions being a product of local velocity and curvature. Another reason to choose \VMD\ as the quantity to define expansions is that many patches of high activity in the local motion kymograph contain multiple \VMD~areas. The \VMD\ allowed us to divide these patches of high local motion into single separated expansions with high \VMD\ rate.

% --------------------------------------------------------------------------------------------------------------------------------------------------
\subsection*{Virtual marker dispersion allows to identify and characterize expansions}\label{ssec:LD}
% --------------------------------------------------------------------------------------------------------------------------------------------------

\begin{figure}
\begin{center}
\showFigure{
\includegraphics[width=0.9\linewidth]{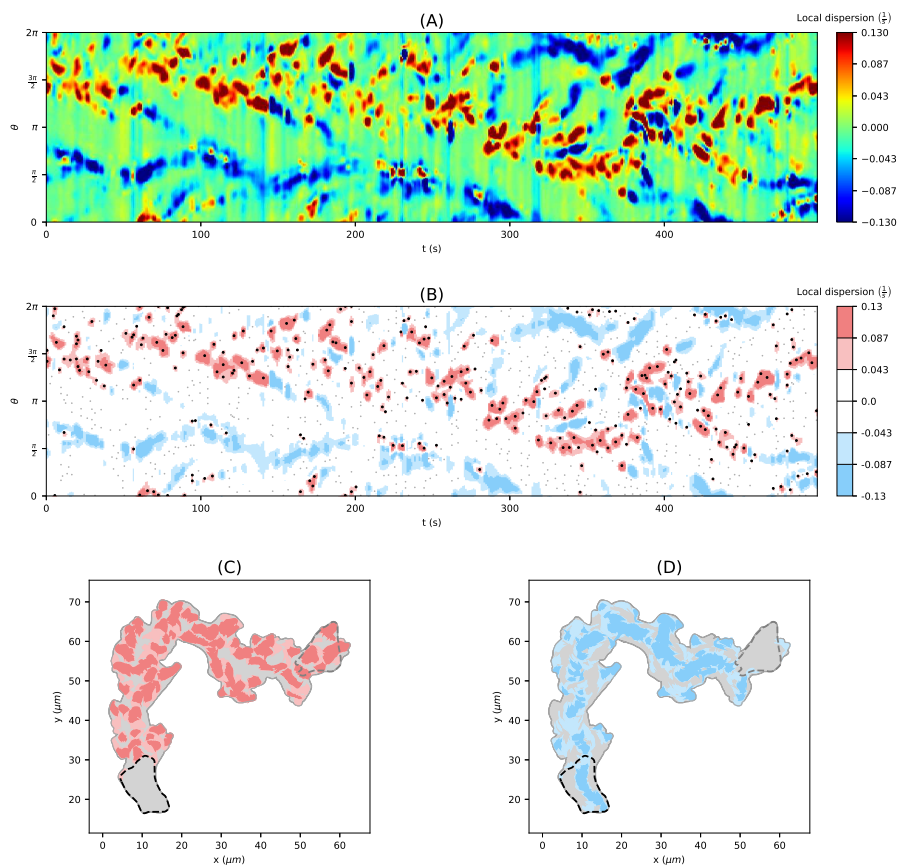}
}
\end{center}
\caption{\textbf{From the local dispersion kymograph to expanding areas and expansion events}. \textbf{(A)} Local dispersion of a persistently motile cell as in Figs~\ref{fig:cell-track-009}~and~\ref{fig:compare_kymo} (left-hand side). \textbf{(B)}
Discretised local dispersion with different areas of activity: high (dark red), medium (light red), low (white) expanding activity, and low (white), medium (light blue), high (dark blue) contracting activity. Local maxima of positive local dispersion are depicted as black dots. Areas of medium and high expanding \textbf{(C)} and contracting \textbf{(D)} activity mapped back on the trace of the cell track.}
\label{fig:feature_identification009}
\end{figure}

Using the persistently motile cell track in Fig~\ref{fig:cell-track-009}, we describe next, how to use the \VMD\ to define expansion areas and expanding events. Based on the local dispersion kymograph in Fig~\ref{fig:feature_identification009}A, we defined areas of medium (light red) and high (dark red) expanding activity as well as medium and high contracting activity (light and dark blue, respectively), see Panel (B). In this context, we determine thresholds for expanding activity by dividing the $90$th percentile of all positive \VMD\ values of the kymograph into three intervals of equal length. For contracting activities ($\VMD<0$) the opposite thresholds were taken. By leaving out the largest and smallest values of the \VMD\ kymograph, the classification becomes less dependent on outliers.
Additionally, we performed a prior smoothing of the kymograph as mentioned in the first paragraph of the previous section to reduce noise and, therefore, to reduce the number of small and separated patterns.

To highlight the events of the highest local expanding activity, we included positive local maxima of the \VMD\ kymograph (black dots) in panel (B). Local maxima falling inside regions with high or middle expanding activity were depicted as bold dots.
Using the time-evolving coordinate system obtained from strongly regularized flow, the expansion/contraction areas shown in panel (B) are mapped back into the 2d plane of amoeboid motion, see panels (C) and (D), respectively. As a result, we obtain an automated visualization of the expanding activity during amoeboid locomotion. 

The kymograph in panel (B) clearly shows the trace of expanding activity at the leading edge, located initially at around $3\pi/2$ and then shifting towards $\pi$. At the same time, contracting activity occurs mainly at a distance of $\pi$ from the leading edge.
In panels (C) and (D), one can nicely see the explorative dynamics of the expansions at the cell front and the stably retracting back of the cell, the so-called uropod, where dark blue areas indicate faster contractions. Analogous graphics for a collection of motile and stationary cells can be found in \hyperref[s:collection-motile]{S13}~and~\hyperref[s:collection-stationary]{S14}~Figs, each containing 12 cell tracks.
\customspace

\begin{figure}
\begin{center}
\showFigure{
\includegraphics[width=0.9\linewidth]{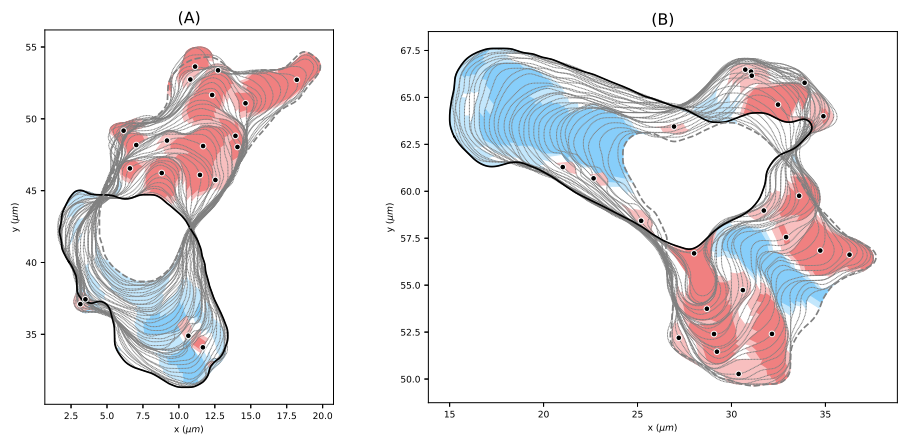}
}
\end{center}
\caption{\textbf{Expanding and contracting areas with corresponding expansion events.} Illustrative sequence of the contour dynamics for $96s\leq t\leq144s$ (left), and  $316s\leq t \leq350s$ (right) based on the cell track shown in Fig~\ref{fig:feature_identification009}. 
Features with high and medium expanding activities are shown in dark and light red, respectively. Features with high and medium contracting activities are shown in dark and light blue, respectively. All patterns shown possess a minimal growth time of $3s$.
The black dots show local maxima of the local dispersion in areas of medium and high expanding activity.
}
\label{fig:shapesOfexpansions}
\end{figure}

In Fig~\ref{fig:shapesOfexpansions}, we present a close-up of two sequences of cell contours. The core expanding areas that shape the evolving cell contour can be seen clearly, e.g., in panel (A). The events of the highest expanding activity (black dots) seem to drive the expansion in many cases. Panel (B) illustrates strong contracting activity at the uropod, and the contraction of the membrane between two nearby expansions (blue area sandwiched between red areas). The opposite effect can be seen in panel (A) at the back of the cell, where a concave region between two convex contractions is identified as expansion. These patterns nicely illustrate the concept of local dispersion, being a product of the local velocity of virtual markers and the curvature along the contour segment.

In \nameref{s:collection}, every identified pattern with minimal growth time $\Delta t\geq 3$ is shown for the underlying cell track.

% --------------------------------------------------------------------------------------------------------------------------------------------------
\subsection*{Statistical analysis of motility patterns}\label{ssec:Statistics}
% --------------------------------------------------------------------------------------------------------------------------------------------------

In this section, we illustrate the ability to statistically analyze a cell track based on our regularized contour flow approach. We used the persistently motile cell track for illustration. 
\customspace

\begin{figure}
\begin{center}
\showFigure{
\includegraphics[width=0.9\linewidth]{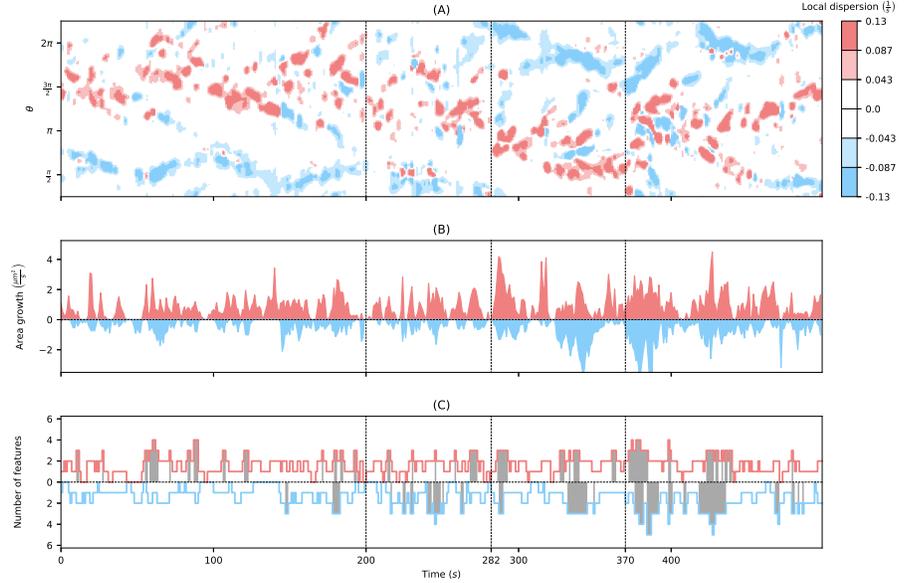}
}
\end{center}
\caption{\textbf{Statistical analysis of example cell track.}
\textbf{(A)} Local dispersion kymograph with thresholds as in Fig~\ref{fig:feature_identification009}.
\textbf{(B)} Area growth of cell segments which are part of identified expansions (red) and contractions (blue) of high and middle intensity.
\textbf{(C)} Number of expanding and contracting areas with high intensity with respect to time. The time with $>2$ features are colored gray to highlight the change in activity in the different phases.  
}
\label{fig:009_stats1}
\end{figure}

Fig~\ref{fig:009_stats1}A shows the local dispersion kymograph of the persistently motile cell divided into four different phases (note that here the y axis starts at $\pi$/4). Until $t=200s$, the cell moves upward with well observable expansions roughly between $1.1\pi$ to $0.2\pi$, while the uropod is slightly above $\pi/2$. Then, the cell begins a phase of reorientation that lasts until $t\approx 280s$, where larger expansions occur also at the former back of the cell. Subsequently, the cell changes its direction downward toward $3/4\pi$ to $\pi$ (in the video it moves rightwards). In the last phase, the cell moves to the right-hand side by creating expansions at the front left, front right, and again front left of the cell. See \nameref{s:009-vid} for a better understanding of the cell track. 
\customspace

We analyzed the expanding and contracting areas shown in Fig~\ref{fig:feature_identification009}C~and~\ref{fig:feature_identification009}D with respect to the activity level (medium/high), duration, and position along the cell contour. In addition, we investigated the differences between expansions and contractions. 
Fig~\ref{fig:009_stats1}B displays the area growth of expansions and contractions for high activity. The identified expanding and contracting areas are naturally partitioning by the sequence of contours into smaller 'slices' (see, e.g., Fig~\ref{fig:shapesOfexpansions}). We defined the area growth of an expansion/contraction as the area of the slice divided by the frame rate $\delta t$. We observed that the overall change of cell area attributed to expansions of high activity is substantially larger than for contractions of high activity. This illustrates that the cell motility of this cell track is driven by a higher number of fast (and potentially explorative) expansions, and a small number of fast contractions. Since in broad terms, the total area gain balances the total area loss, it further illustrates that area loss is to a larger extent attributed to slower and steadier contractions than it is for expansions.  After the second phase ($t>280$), when the cell has completed reorientation, the area change for both, expansions and contractions increased and even more so at the beginning of the last phase ($t>370$). 

Finally, in panel (C), the number of simultaneous expansions and contractions is shown (high activity only). In the first phase of the upward moving cell, the number of expansions is much higher than the number of contractions (mean: $1.52$ vs.\ $1.12$). In addition, we determined the fraction of time $fT_{\text{exp}>2}$ and $fT_{\text{contr}>2}$ with more than two simultaneous expansions and contractions, respectively.
In the first phase, it is $fT_{\text{exp}>2}= 0.146$ vs.\ $fT_{\text{contr}>2}=0.035$.
In the following phases, these fractions increase for contractions: $fT_{\text{contr}>2}= 0.198; 0.195; 0.341$ (for phase $2; 3; 4$), whereas for expansions the fraction is slightly smaller in the second phase and larger in the last two phases:
$fT_{\text{exp}>2}= 0.123; 0.207; 0.217$ (for phase $2;3;4$).
This illustrates that the increased activity, as seen in panel (B), goes along with an increased number of expansions and contractions. Moreover, this indicates a more explorative character of the cell motion in phases three and four, while the cell seemed to be more stabilized during the first phase.
\customspace

\begin{figure}
\begin{center}
\showFigure{
\includegraphics[width=0.85\linewidth]{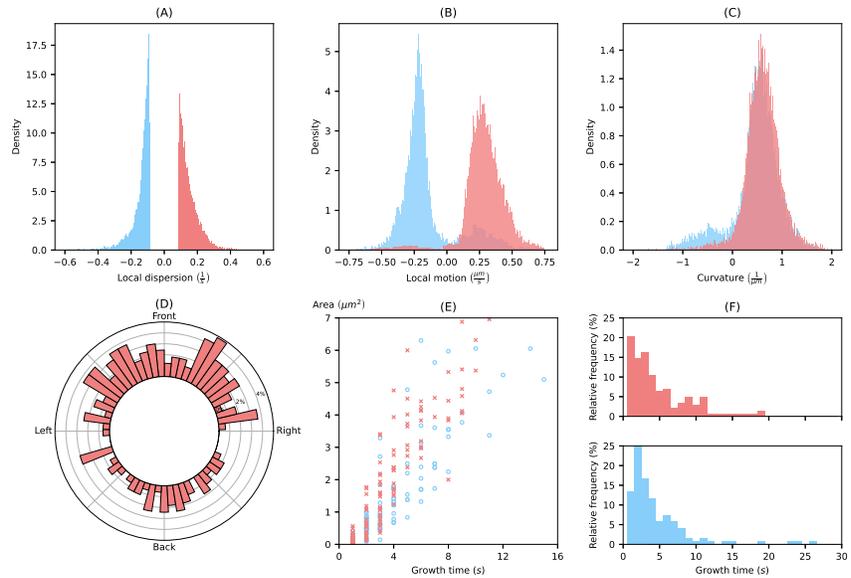}
}
\end{center}
\caption{\textbf{Statistical analysis of example cell track.}
Distributions of local dispersion \textbf{(A)},
local motion \textbf{(B)} and curvature \textbf{(C)}  inside expanding and contracting patterns with high intensity.
\textbf{(D)} Circular histogram displaying the angle, where high expanding activity appears along the cell contour.
\textbf{(E)} Correlation between area and growth time of identified patterns.
\textbf{(F)} Histograms of growth times of expansions and contractions with high intensity.
}
\label{fig:009_stats2}
\end{figure}

Fig~\ref{fig:009_stats2} gives further insight into the expanding activity.
Panel (A) shows the distribution of \VMD\ values of all virtual markers within expanding and contracting areas with high activity. In other words, the local dispersion distribution shows only values that are larger than the thresholds for high expanding activity or smaller than the threshold for high contracting activity (the thresholds are $\pm 0.087$). The distribution of the \VMD\ of expansions further extends towards large values than the corresponding distribution of contraction towards small values (median of $0.13$/s vs.\ $-0.12$/s). 

In panels (B) and (C), the distributions of local motion and the curvature are shown for all virtual markers within expanding and contracting areas of high activity (same areas as above). 
For the local motion, we observed major peaks around $0.29\,\mu m/s$ and $-0.21\,\mu m/s$ for expansions and contractions, respectively. Minor peaks correspond to inward expansions and outward contractions discussed in the previous section and shown in Fig~\ref{fig:shapesOfexpansions} (see red expanding areas within a blue contracting region in panel (A), and a blue contracting area within an expanding region in panel (B)). In line with this observation, minor peaks of concave (negative) curvature and major peaks for convex (positive) curvature are shown.

Fig~\ref{fig:009_stats2}D shows the distribution of high activity expansions in direction of the moving cell. We identified two peaks in direction of the cell movement (front-left and front-right). Another peak is located at the back of the cell. A similar behaviour was presented for pseudopods in~\cite{bosgraaf_ordered_2009,bosgraaf_navigation_2009}, where two different types of pseudopods were distinguished: (left/right) splitting pseudopods and \textit{de-novo} pseudopods.
By comparing the correlation between the growth time and the area of expansions and contractions in panel (E), we observed that expansions possess an area often twice as large as contractions with similar growth times. This indicates the difference between the faster and more explorative character of expansions at the cell front and the slower and stably retracting character of the uropod. This is also in line with the corresponding activities shown in Fig~\ref{fig:009_stats1}B.

Finally, in panel (F) the distribution of growth times is shown for both, high activity expansions and contractions. Note that we used the term 'growth time' for both, expansions as well as contractions, owing to the fact that also contracting areas can be moving in an outward direction, as discussed earlier. 
The majority of growth times fall inside a range of $0s$ to $10s$. Nevertheless, there are patterns, especially long persistent uropods, with growth times much larger than the range presented in these histograms. The average growth time of pseudopods observed in \cite{bosgraaf_ordered_2009} is much higher ($12.8\,s$) than the growth times of expansions presented in this work ($4.9\,s$).
This is not surprising, since our definition of expansions also takes short-lived objects into account. For example, the average number of expansions per minute for the persistently motile cell track ($\approx 17.0$) is much higher than the average frequency of pseudopods per minute ($2.9\pm 0.2$) as observed in \cite{bosgraaf_ordered_2009}.
Additionally, concave regions between two contractions as in Fig~\ref{fig:shapesOfexpansions}A were detected as expanding areas as well, which raises the total number of expansions.
See \hyperref[s:collection-motile]{S13}~and~\hyperref[s:collection-stationary]{S14}~Figs for similar graphics of
a collection of 24 cell tracks, also containing the other two tracks from Fig~\ref{fig:compare_kymo}.

\subsection*{Application to other kinds of cell motility}
In this section, we demonstrate the flexibility of our method to study other kinds of cell motility. To this end, 
we extracted sequences of cell contours from videos of early embryonic killifish cells (\textit{Fundulus heteroclitus}) \cite{Fink_7792, Fink_35204, Fink_35205} and keratocytes cultured from
Central American cichlids (\textit{Hypsophrys nicaraguensis}) \cite[S1-S3 Movies]{Barnhart2011}. While the images of the embryonic killifish cells were obtained from fluorescence microscopy, bright-field microscopy was used to record the keratocytes. As before, the images were segmented by using a modified version of the active contour (snake) algorithm described in~\cite{driscoll_local_2011,driscoll_cell_2012}.

\begin{figure}
\begin{adjustwidth}{-2.25in}{0in}
\begin{center}
\showFigure{
\includegraphics[width=0.85\linewidth]{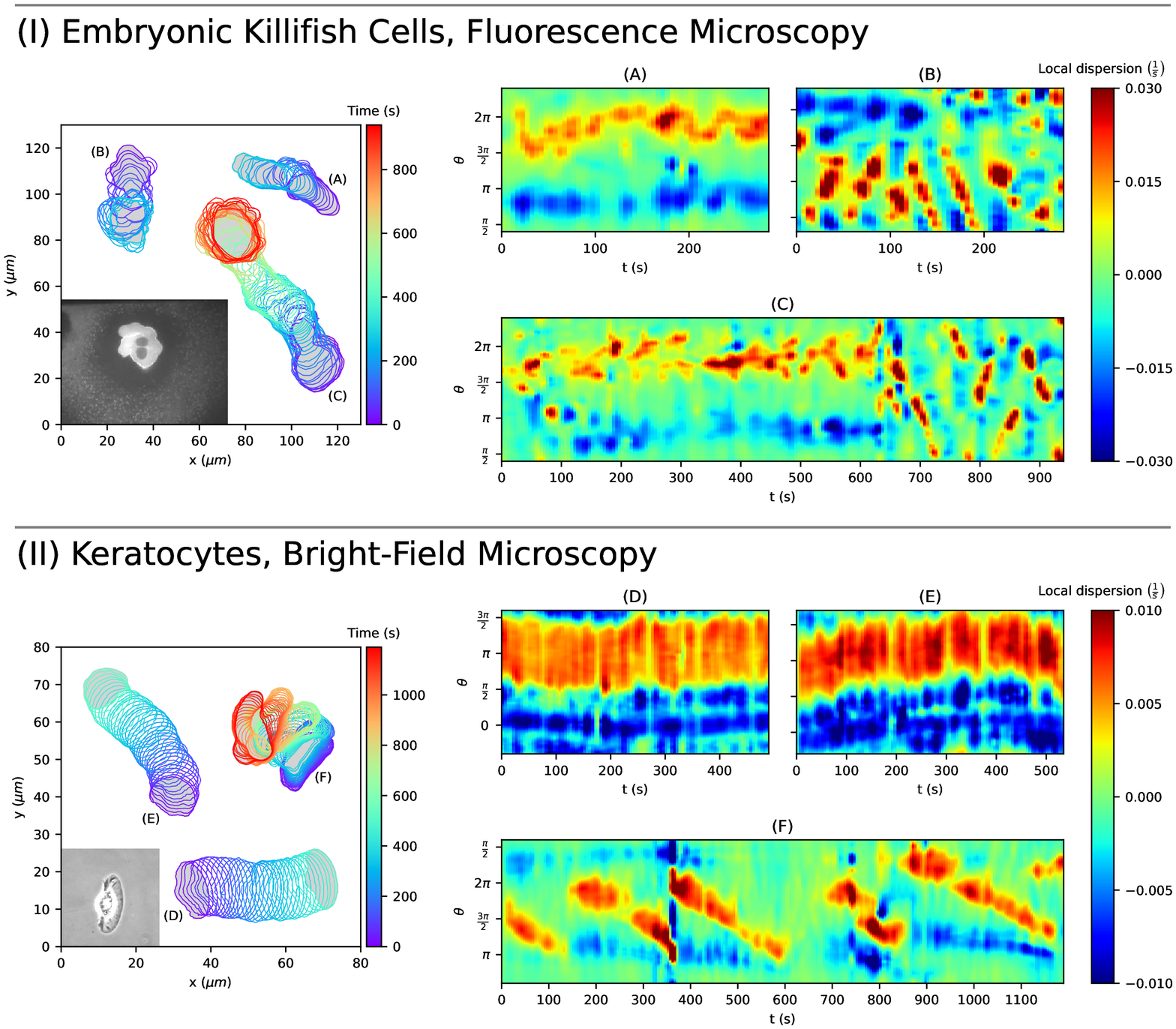}
}
\end{center}
\caption{\textbf{Application to other kinds of cell motility.}
Displayed are three tracks for embryonic killifish cells \textbf{(A-C)} and keratocytes \textbf{(D-E)}, and their corresponding local dispersion kymographs. Protrusive migration can be seen in \textbf{(A, C)}, circular waves around the cell border in \textbf{(B, C, F)}, and a steady persistent translation of the contour in \textbf{(D, E)}.
}
\label{fig:other_data}
\end{adjustwidth}
\end{figure}

In Fig~\ref{fig:other_data}, three cell tracks for both applications, embryonic killifish cells and keratocytes, and corresponding local dispersion kymographs are shown. In panels (A) and (C), protrusion-driven cell motility can be observed for the embryonic killifish cells. Moreover, rotating waves around the leading edge (so-called circular movements) can be seen in panel (B) and in the last third of panel (C) as diagonal lines. In the chosen parametrization, right upward diagonals indicate counterclockwise rotations of protrusions while right downward diagonals indicate clockwise rotations. Notably, the switching between these kinds of locomotion can be nicely seen in panel (C) at $t\approx 630$. In addition, we refer to the video accessible on \cite{Fink_35204} showing the cell track from panel (B) with dominant circular movements.

On the contrary, keratocytes possess a steady and persistent type of cell migration with only minor shape deformations. However, keratocytes can also show a more crawling kind of cell motility depending on the adhesion strength. In \cite[S1-S3 Movies]{Barnhart2011}, cell track recordings for different adhesion strengths are provided: intermediate adhesion strength (panel (D)), low adhesion strength (panel (E)), and high adhesion strength (panel (F)). For low and intermediate adhesion strength, persistent cell track can be observed with a slightly higher local dispersion at the cell front for the case of low adhesion strength.
For high adhesion strength, the cell migrates slower and exhibits larger cell shape deformations. Again, diagonal lines in panel (F) indicate circular movements as described earlier. Due to segmentation/mapping difficulties resulting from jumps in the video recording, pronounced artifacts can be seen as vertical lines at $t=360s$ and $t=750s$.

These test applications underline the flexibility of our algorithm, handling contours obtained from fluorescence images as well as bright-field recordings. Moreover, the underlying data was recorded with a lower temporal resolution, $5s$ per image compared to only $1s$ in our data.
For the case of embryonic killifish cells, which possess local contour changes comparable to our data sets, a temporal resolution of $5s$ is relatively low. This can be nicely seen in panel (B).
Since the movement of keratocytes is much slower with less prominent contour changes, kymographs with relatively high resolution can be acquired for a sufficient temporal resolution of $5s$.
This shows the feasibility of our algorithm for different imaging frequencies, see also \nameref{s:temporalresolution}.
Finally, the local dispersion kymographs of these test cases show distinct differences from previous kymographs regarding amoeboid cell migration. This indicates the potential of our algorithm for classifying different kinds of cell motility.

% ===================================================================================
\section*{Discussion}
% ===================================================================================

With the ever-increasing amount of live cell imaging data and the continuously growing computational power, computer-automated techniques to analyze the morphology of cells have steadily developed over the past two decades. 
In particular, in cell motility research, morphological characteristics are commonly used to pinpoint phenotypic differences between mutant cell lines thus highlighting the mechanistic role of individual components of the underlying signaling pathways. 
While many static measures of cell shape have already been introduced early on~\cite{murray_three-dimensional_1992,wessels_computer-assisted_1998}, dynamic measures that quantify the temporal evolution of the cell shape proved to be more difficult to implement.
First attempts focused on temporal changes of the projected cell area to deduce overall protrusion rates, for an example see~\cite{dunn_dynamics_1995}.
These approaches were later refined by local measures of cell boundary motion along selected line segments perpendicular to the cell border~\cite{rottner_vasp_1999}. 
Also, local space-time plots have been defined in this way~\cite{hinz_quantifying_1999,totsukawa_distinct_2004}. 
However, in all these cases, the direction of interest or the local placement of the kymograph had to be chosen manually, which severely limits a reliable long-time tracking of more complex cell shapes and introduces an arbitrariness related to the manual processing.

The most promising approach to overcome this limitation relies on an active contour (snake), a closed chain of connected nodes (virtual markers) that is placed along the cell perimeter~\cite{dormann_simultaneous_2002}.
Different rules have been proposed to propagate the markers from one contour to the next.
In some cases, the distance between virtual markers is kept constant and markers are added or removed as the contour evolves~\cite{bosgraaf_analysis_2009}.
Other approaches that are inspired by mechanical spring models or concepts from electrostatics keep the number of markers constant and allow for local variations in the marker spacing~\cite{tyson_high_2010}.
Here, we can distinguish two limiting cases. 
On the one hand, equidistance is enforced and markers on adjacent contours are connected in a one-to-one mapping by minimizing the sum of square distances between pairs of connected markers~\cite{driscoll_local_2011}.
On the other hand, markers are propagated from one contour to the next in a normal direction (normal flow) while the marker spacing evolves without constraints.
The latter approach has been implemented using a level set method to cope with problems related to finite time sampling~\cite{machacek_morphodynamic_2006}.
However, high computational costs and the rapid buildup of highly uneven marker distributions limit the use of the level set method in practical applications.

In this article, we have introduced a family of marker flows that incorporates these different cases into a general framework.
In particular, the regularization that we introduced in Eq~\eqref{eq:regularization} includes the two extreme scenarios described above as limiting cases.
In the limit of large $\lambda$, we obtain an equidistant mapping, whereas, in the limit of small $\lambda$, we approach the shortest path flow, respectively, the reverse normal flow.
Tuning $\lambda$ allows us to systematically shift between these two limits.

Once a flow of virtual markers is computed on the evolving cell contour by any of these methods, it defines a coordinate system, in which different local quantities can be displayed, such as curvature, membrane displacement, or the intensity of a fluorescently tagged membrane-associated protein. 
%see Ref.~\cite{baniukiewicz_quimp:_2018} for a specific software tool, where some of this has been implemented.
%
Essential to our approach is the clear separation of local quantities derived by weakly regularized flows between two consecutive contours and the coordinate system which is based on a single strongly regularized (global) flow onto which each quantity of interest is mapped. This way, trajectories of each quantity can be obtained for the entire time period.
Note that for all of these coordinate choices it is generally acknowledged that the dynamics of the virtual membrane markers do not reflect the motion of specific membrane lipids or proteins, as the membrane itself is a very complex and dynamic structure~\cite{machacek_morphodynamic_2006,driscoll_local_2011}.
In particular, lateral flows may occur due to membrane recycling, so that the dynamics of individual molecules or domains in the membrane do not necessarily correspond to morphological changes and will prevent a one-to-one mapping of molecular markers on adjacent contours.
\customspace

Amoeboid motion is primarily driven by localized membrane protrusions, so-called pseudopodia.
Identifying and tracking pseudopodia has thus been an important focus of the morphodynamic analysis of amoeboid cells.
The first substantial pseudopod statistics were generated by computer-assisted manual image processing, relying on the expert judgment of the investigator~\cite{andrew_chemotaxis_2007,bosgraaf_ordered_2009}.
From this, a first automated software routine for pseudopod tracking was developed~\cite{bosgraaf_quimp3_2010} and successfully applied also to analyze the chemotactic navigation of amoeboid cells~\cite{bosgraaf_navigation_2009}.
It relies on a complex decision tree that defines pseudopodia based on a sequence of threshold criteria applied to the local curvature, the virtual marker movement, and the local area change.
In this way, the frequency and direction of pseudopod formation, their sizes, lifetimes, and other quantitative measures were extracted.
While successfully providing a first quantitative database of pseudopod characteristics, this approach has the drawback that it requires the choice of several parameters that are tuned to the characteristic properties of pseudopods in starvation-developed {\it D.~discoideum} cells.
If cells display protrusions with a more diverse range of shapes and time scales, a reliable tracking is difficult to achieve with this approach.

Later, a more compact criterion for the detection of localized protrusions was proposed~\cite{driscoll_local_2011,driscoll_cell_2012}.
It relies on thresholding a distance measure between the current position of virtual markers and the cell boundary at a later time point.
The calculation of this distance measure, however, lacks a clear underlying definition and is computed in an ad hoc fashion for the specific data set (see supplementary material of Ref.~\cite{driscoll_cell_2012}):
First, each virtual marker is mapped from its current position onto the closest point on the future cell contour.
As this generates a highly non-uniform distribution of target markers, with protrusive areas particularly poorly covered, the target markers on the new contour are then redistributed by two successive smoothing steps, using averaging windows of specific sizes.
The time interval between the successive contours for the distance projection, as well as the smoothing parameters for redistribution of the target markers, were hand-picked by the investigator.
Note, however, that this could be envisioned as one step in an iterative procedure to minimize our cost functional.

In the present work, we introduce a novel approach to define, identify and analyze localized expansions on dynamically evolving contours of amoeboid cells.
expanding areas are defined via a single threshold value. 
In contrast to previous approaches, we chose the virtual marker dispersion as the underlying quantity, since it combines information on both, the marker displacement and the local curvature. The marker dispersion is mathematically well defined by Eq~\eqref{eq:reinitializedlocaldispersion}  and does not require additional empirical smoothing steps. 
Based on this criterion, we not only detect individual expansion events but we capture the entire shape and the complete temporal evolution of an expansion in a fully automated fashion, see Fig~\ref{fig:shapesOfexpansions}~and~\hyperref[s:collection]{S15}~Fig for example.

An implementation of the methodology, obtaining kymographs from regularized flows in general and detecting expansion/contraction patterns from these kymographs in particular, are fully accessible and well documented in our software Package \texttt{AmoePy}. We also provide the data of multiple cell tracks and simple artificial test cases on which the algorithm was validated.
Finally, all figures and results from this article can be easily reproduced in \texttt{AmoePy}.

\begin{figure}
\begin{center}
\showFigure{
\includegraphics[width=0.85\linewidth]{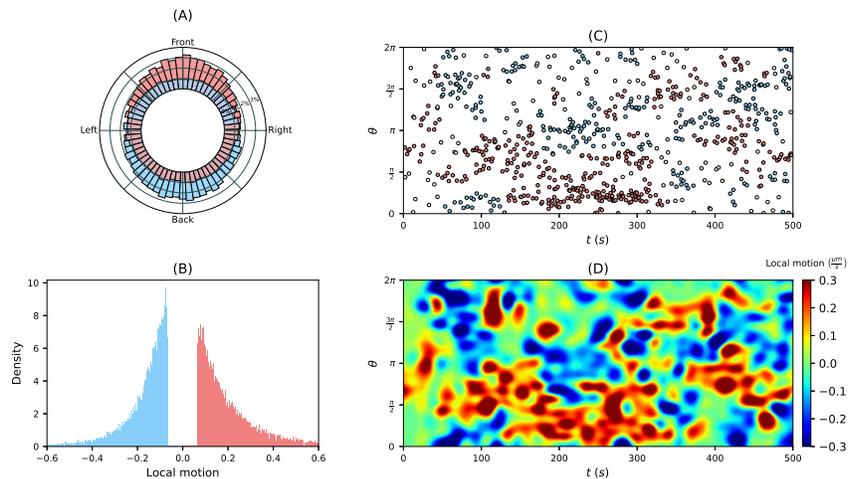}
}
\end{center}
\caption{\textbf{Future outlook.}
Underlying distributions are based on averaging over several cell tracks. 
\textbf{(A)} Circular histogram displaying the position, where expansions and contractions events appear along the cell contour.
\textbf{(B)} Distribution of local motion of expansion and contractions events above a threshold $\pm 1/3$.
\textbf{(C)} Simulated data obtained from self-excited Poisson point processes (so-called Hawkes processes) on the unit circle.
Afterward, we obtained regions of high (red) and low (blue) intensity w.r.t. to a clustering algorithm.
\textbf{(D)} Continuous kymograph obtained from a regression model (e.g. Gaussian process regression) based on sampled magnitudes at event locations shown in (C) from the distribution in (B). 
}
\label{fig:outlook}
\end{figure}

As outlined in the Introduction, the overall aim is a quantitative, data-driven model of amoeboid motility. The presented theoretical framework is a first step in this direction. Because of its rigorous mathematical formulation, its efficient avoidance of mapping violations during larger shape deformations, and its moderate computational costs, it is a suitable choice for such a model.
We envision that point events of high expanding activity may be used to define a point process in the space-time coordinate system. To reflect the often observed persistence in motility, so-called self-excited Poisson cluster processes or Hawkes processes may be favorable choices. The point process can serve as a skeleton for expansion activity that is 'completed' to a random realization of a kymograph, based on the statistics of a local quantity. We illustrated this idea in Fig~\ref{fig:outlook}, where we used the local motion statistics (B) to reconstruct a local motion kymograph (D). The idea is to use such realizations of kymographs to reconstruct a cell track and eventually to assimilate the time-lapse microscopy data into a mathematical model of amoeboid motility.

%
%\paragraph*{S1 Table.}
%\label{S1_Table}
%{\bf Lorem ipsum.} Maecenas convallis mauris sit amet sem ultrices gravida. Etiam eget sapien nibh. Sed ac ipsum eget enim egestas ullamcorper nec euismod ligula. Curabitur fringilla pulvinar lectus consectetur pellentesque.

% ===================================================================================
\section*{Acknowledgments}
% ===================================================================================
%This research has been partially funded by Deutsche Forschungsgemeinschaft (DFG) - SFB1294/1 - 318763901.
We thank Zahra Eidi (IPM, Tehran/Iran), Gregor Pasemann (Technische Universit\"at Berlin), and Till Bretschneider \& Piotr Baniukiewicz (University of Warwick/UK) for fruitful and stimulating discussions. We also thank Lena Lindenmeier for additional help regarding the graphical user interface and testing routines of \texttt{AmoePy}.

\nolinenumbers

% Either type in your references using
% \begin{thebibliography}{}
% Text
% \end{thebibliography}
%
% or
%
% Compile your BiBTeX database using our plos2015.bst
% style file and paste the contents of your .bbl file
% here. See http://journals.plos.org/plosone/s/latex for 
\section*{Supporting information}
% ===================================================================================

% Include only the SI item label in the paragraph heading. Use the \nameref{label} command to cite SI items in the text.

\paragraph*{S1 Text.}
{\bf Supplementary methods and supporting computations.} (PDF)
\label{s:appendix1}

\paragraph*{S1 Fig.}
\label{s:amoebatube}
{\bf Three-dimensional space--time tube of contours}, where consecutive contours are stacked onto each other. (PDF)

\paragraph*{S2 Fig.}
\label{s:compareregularization}
{\bf Comparison of virtual marker mappings obtained with different regularization schemes}, using the distances between neighboring virtual markers in $\mathbb{R}^2$ and $\mathcal{S}^1$, respectively. (PDF)

\paragraph*{S3 Fig.}
\label{s:largeshapedeformations}
{\bf Selection of contour mappings under large shape deformations} with corresponding local motion kymograph.
The underlying cell track was recorded with an imaging rate of $\delta t\approx 3.13s$. (PDF)

\paragraph*{S4 Fig.}
\label{s:testcases}
{\bf Collection of 13 artificial cell tracks} with corresponding kymographs showing local dispersion, local motion, and curvature.
In the figure at the top of each page, the period of time of several cell tracks was shortened for illustrative purposes due to overlapping contours. (PDF)

\paragraph*{S5 Fig.}
\label{s:pseudocodeMF}
{\bf Schematic overview of an algorithm} to obtain coordinate markers from solving the optimization problem as in Eq~\eqref{eq:reinitregularization}. See also \nameref{s:appendix1} for more details. (EPS)

\paragraph*{S6 Fig.}
\label{s:comptime}
{\bf Computation times of ``RegFlow'' algorithm shown in Fig~\ref{fig:pseudocodeRF}} for different regularization parameters and contours. (PDF)

\paragraph*{S7 Fig.}
\label{s:contourestimate}
{\bf Cell contour based on noisy fluorescence images} estimated via GPR for different hyperparameters and corresponding curvature. (PDF)

\paragraph*{S8 Fig.}
\label{s:temporalresolution}
{\bf Comparison of local motion kymographs obtained with different imaging frequencies:} $\delta t\in\{1,2,3,5,10\}$. (PDF)

\paragraph*{S9 Fig.}
\label{s:globalflow}
{\bf Comparison of global flows} in cases of no, weak and strong regularization. (PDF)

\paragraph*{S10 Fig.}
\label{s:globalflowtestcase}
{\bf Test scenario of a cell track}, where the underlying coordinate system is based on only 8 out of 500 contours. (PDF)

\paragraph*{S11 Fig.}
\label{s:compkymo}
{\bf Kymographs as in Fig~\ref{fig:compare_kymo} without prior smoothing.} (PDF)

\paragraph*{S12 Fig.}
\label{s:correlation}
{\bf Correlation between local dispersion and local motion.}
The correlation is shown for the cell track in Fig~\ref{fig:compare_kymo}: Persistently motile, weakly motile and almost stationary cell. (PDF)

\paragraph*{S13 Fig.}
\label{s:collection-motile}
{\bf Collection of 12 cells with high and medium motility} and different grades of persistence. For each track the kymographs of local dispersion, local motion and curvature are shown, followed by plots as in Fig~\ref{fig:009_stats1} and~\ref{fig:009_stats2}.
The cells are sorted in descending order regarding the area of the entire contour track. (PDF)

\paragraph*{S14 Fig.}
\label{s:collection-stationary}
{\bf Collection of 12 stationary cells.} The file is structured as in \nameref{s:collection-motile}.
(PDF)

\paragraph*{S15 Fig.}
\label{s:collection}
{\bf Collection of identified expanding areas and events} of high intensity for the persistently motile cell. Only features with minimal persistence length $\Delta t\geq 3$ are shown. (PDF)

\paragraph*{S1 Video.}
\label{s:009-vid}
{\bf Persistently motile cell.} Cell tracks in all videos are shown at tenfold speed. (MP4)

\paragraph*{S2 Video.}
\label{s:058-vid}
{\bf Weakly motile cell.} (MP4)

\paragraph*{S3 Video.}
\label{s:018-vid}
{\bf Stationary cell.} (MP4)

%\bibliography{2020_schindler_plos_article.bib}
%\nocite{*}

% ==================================================================================
\end{document}